\documentclass[11pt]{article}
\pdfoutput=1
\usepackage{amsfonts,amsmath}
\usepackage{amssymb}
\usepackage{fancyhdr}
\usepackage{slashed}
\usepackage{graphicx}
\usepackage{subfigure}
\usepackage{color}
\usepackage{cite}

\usepackage{hyperref}
\usepackage[utf8]{inputenc}
\usepackage[titletoc]{appendix}

\def\hybrid{\topmargin -20pt    \oddsidemargin 0pt
	\headheight 0pt \headsep 0pt
	\textwidth 6.25in       % A4 paper
	\textheight 9.25in       % A4 paper
	\marginparwidth .875in
	\parskip 5pt plus 1pt   \jot = 1.5ex}

\hybrid

\def\baselinestretch{1.2}

\catcode`\@=11

\def\marginnote#1{}
\newcount\hour
\newcount\minute
\newtoks\amorpm
\hour=\time\divide\hour by60
\minute=\time{\multiply\hour by60 \global\advance\minute by-\hour}
\edef\standardtime{{\ifnum\hour<12 \global\amorpm={am}%
		\else\global\amorpm={pm}\advance\hour by-12 \fi
		\ifnum\hour=0 \hour=12 \fi
		\number\hour:\ifnum\minute<10 0\fi\number\minute\the\amorpm}}
\edef\militarytime{\number\hour:\ifnum\minute<10 0\fi\number\minute}
\def\draftlabel#1{{\@bsphack\if@filesw {\let\thepage\relax
			\xdef\@gtempa{\write\@auxout{\string
					\newlabel{#1}{{\@currentlabel}{\thepage}}}}}\@gtempa
		\if@nobreak \ifvmode\nobreak\fi\fi\fi\@esphack}
	\gdef\@eqnlabel{#1}}
\def\@eqnlabel{}
\def\@vacuum{}
\def\draftmarginnote#1{\marginpar{\raggedright\scriptsize\tt#1}}

\def\draft{\oddsidemargin -.5truein
		\def\@oddfoot{\sl preliminary draft \hfil
			\rm\thepage\hfil\sl\today\quad\militarytime}
		\let\@evenfoot\@oddfoot \overfullrule 3pt
		\let\label=\draftlabel
		\let\marginnote=\draftmarginnote
		\def\@eqnnum{(\theequation)\rlap{\kern\marginparsep\tt\@eqnlabel}%
			\global\let\@eqnlabel\@vacuum}  }
	
	\def\preprint{\twocolumn\sloppy\flushbottom\parindent 2em
		\leftmargini 2em\leftmarginv .5em\leftmarginvi .5em
		\oddsidemargin -.5in    \evensidemargin -.5in
		\columnsep .4in \footheight 0pt
		\textwidth 10.in        \topmargin  -.4in
		\headheight 12pt \topskip .4in
		\textheight 6.9in \footskip 0pt
		\def\@oddhead{\thepage\hfil\addtocounter{page}{1}\thepage}
		\let\@evenhead\@oddhead \def\@oddfoot{} \def\@evenfoot{} }
	
	\def\numberbysection{\@addtoreset{equation}{section}
		\def\theequation{\thesection.\arabic{equation}}}
	
	\def\underline#1{\relax\ifmmode\@@underline#1\else
		$\@@underline{\hbox{#1}}$\relax\fi}

	\def\titlepage{\@restonecolfalse\if@twocolumn\@restonecoltrue\onecolumn
		\else \newpage \fi \thispagestyle{empty}\c@page\z@
		\def\thefootnote{\fnsymbol{footnote}} }
	
	\def\endtitlepage{\if@restonecol\twocolumn \else \newpage \fi
		\def\thefootnote{\arabic{footnote}}
		\setcounter{footnote}{0}}  %\c@footnote\z@ }

\catcode`@=12
\relax

\def\figcap{\section*{Figure Captions\markboth
		{FIGURECAPTIONS}{FIGURECAPTIONS}}\list
	{Figure \arabic{enumi}:\hfill}{\settowidth\labelwidth{Figure
			999:}
		\leftmargin\labelwidth
		\advance\leftmargin\labelsep\usecounter{enumi}}}
 \relax
\def\tablecap{\section*{Table Captions\markboth
		{TABLECAPTIONS}{TABLECAPTIONS}}\list
	{Table \arabic{enumi}:\hfill}{\settowidth\labelwidth{Table
			999:}
		\leftmargin\labelwidth
		\advance\leftmargin\labelsep\usecounter{enumi}}}
 \relax
\def\reflist{\section*{References\markboth
		{REFLIST}{REFLIST}}\list
	{[\arabic{enumi}]\hfill}{\settowidth\labelwidth{[999]}
		\leftmargin\labelwidth
		\advance\leftmargin\labelsep\usecounter{enumi}}}
 \relax
	\makeatletter
	\newcounter{pubctr}
	\def\publist{\@ifnextchar[{\@publist}{\@@publist}}
	\def\@publist[#1]{\list
		{[\arabic{pubctr}]\hfill}{\settowidth\labelwidth{[999]}
			\leftmargin\labelwidth
			\advance\leftmargin\labelsep
			\@nmbrlisttrue\def\@listctr{pubctr}
			\setcounter{pubctr}{#1}\addtocounter{pubctr}{-1}}}
	\def\@@publist{\list
		{[\arabic{pubctr}]\hfill}{\settowidth\labelwidth{[999]}
			\leftmargin\labelwidth
			\advance\leftmargin\labelsep
			\@nmbrlisttrue\def\@listctr{pubctr}}}
	 \relax
	\makeatother
	\newskip\humongous \humongous=0pt plus 1000pt minus 1000pt

	\newif\ifdtup

	\relax
	
	\def\be{\begin{equation}}
		\def\ee{\end{equation}}
	\def\ba{\begin{eqnarray}}
		\def\ea{\end{eqnarray}}

	\def\k{\kappa}
	\def\r{\rho}
	\def\a{\alpha}

	\def\P{\Pi}

	\def\th{\theta}

	\def\om{\omega}

	\def\s{\sigma}

	\def\cM{{\cal M}}

	 \def\cB{{\cal B}} 
	  \def\cF{{\cal F}}
	 \def\cH{{\cal H}} 
	  
	\def\cM{{\cal M}}  
	  
	 \def\cT{{\cal T}} 
	  
	 \def\cZ{{\cal Z}}

	\newcommand{\prt}[1]{{\left( {#1} \right)}}
	\newcommand{\prtt}[1]{{\left[ {#1} \right]}}

	\def\no{\noindent}
	
	\def\qq{\qquad}

	\def\IR{\relax{\rm I\kern-.18em R}}

	\def\pp{\partial}
	
	\newcommand{\ff}{\frac}

	\def\IR{\relax{\rm I\kern-.18em R}}
	\def\IL{\relax{\rm I\kern-.18em L}}
	
	\def\inv{^{\raise.15ex\hbox{${\scriptscriptstyle -}$}\kern-.05em 1}}

	\def\cM{{\cal M}}

	\def\bea{\begin{eqnarray}}
		\def\eea{\end{eqnarray}}
	\newcommand{\eq}[1]{(\ref{#1})}
	\def\nn{\nonumber}

	\newcommand{\la}[1]{\label{#1}}

	\def\a{\alpha}

	\def\f{\phi}        

	\def\k{\kappa}

	\def\o{\omega}
	 \def\P{\Pi}
	\def\r{\rho}
	\def\s{\sigma}  
	\def\t{\tau}
	\def\th{\theta}

	\definecolor{markcolor2}{rgb}{1,0,0}
	
	\definecolor{markcolor3}{rgb}{0,1,0}

\begin{document}
		
		\begin{titlepage}

			\begin{center}
				
				%\hfill nsysu-xx-yyyy\\
				%\vskip -.1 cm
				%\hfill hep--th/yymmnnn\\
				
				~
				\vskip .9 cm

				{\large
					\bf Holographic Non-local Rotating Observables and their Renormalization }
				
				\vskip 0.4in
				{\bf Vangelis Giantsos$^{1}$, Dimitrios Giataganas$^{*~ 2,3}$ }
				\vskip 0.2in
				{\em
					${}^1$ Department of Physics,\\ University of Athens,  Zographou 157 84, Greece\vskip .1in
					${}^2$  Department of Physics,\\ National Sun Yat-Sen University,  \\
					Kaohsiung 80424, Taiwan
					\vskip .1in
					${}^3$  Center for Theoretical and Computational Physics, \\
					National Sun Yat-Sen University					
					\\
					Kaohsiung 80424, Taiwan
					\\ \vskip .15in
					{\tt ${}^*$ dimitrios.giataganas@mail.nsysu.edu.tw}
				}

				\vskip .1in
			\end{center}
			
			\vskip 1.1in
			
			\centerline{\bf Abstract}
We analyse non-local rotating observables in holography corresponding to spinning bound states. To renormalize their energies and momenta we suggest and discuss different holographic renormalization schemes motivated by the static non-local observables.  Namely the holographic renormalization and the rotating color singlet mass subtraction scheme.  
In the holographic renormalization we identify the infinite boundary terms and subtract them. In the mass subtraction scheme we evaluate the energy of a spinning trailing string corresponding to the color charged singlet which experiences dragging phenomena and we subtract it from the energy of the bound state to obtain the renormalized finite energy.
Then we apply our generic framework to certain strongly coupled thermal theories with broken rotational symmetry. We find numerical solutions corresponding to spinning bound states with a fixed size while varying their angular frequency. By applying numerically the renormalization schemes, we find that there is a critical frequency where the bound state ceases to exist or dissociates. 
We also note that bound states require lower angular frequencies to dissociate when the theory has less symmetry.
			\no
		\end{titlepage}
		\vfill
		\eject

		%\end{center}

		\noindent

		%\vskip .4in
		%\noindent
		%August 2002\\
		%\end{titlepage}
		%\vfill
		%\eject
		
		\def\baselinestretch{1.2}
		\baselineskip 19 pt
		\noindent
		
		%%%%%%%%%%%%%%%
		
		\setcounter{equation}{0}
		
\section{Introduction}

Non-local observables have been studied extensively in the context of gauge/gravity dualities, providing several interesting insights for the duality and the non-perturbative physics of the non-abelian gauge fields. The central example of such observables is the static Wilson loop and the entanglement entropy, where their expectation values are related in holography to extremal surfaces of certain static boundary conditions. The analytical computation of extremal surfaces, in presence of scales of the theory, like the finite temperature, is challenging and usually can be done only with series expansions. At the same time it is natural that it contains richer information about the theory that probes. For example, the Wilson loop expectation value provides information on confinement, the stability and the dissociation of static heavy quark bound states in the dual theory. In the holographic approach, the computation consists of the classical on-shell string action of a worldsheet that is attached on the boundary of the space, where its extremal area diverges. There are several naturally motivated proposals on the renormalization scheme that need to be invoked to cancel this divergence. Each of them can be thought as corresponding eventually to a different type of observable since the finite part of  the resulting expectation value depends on the scheme applied.

The vast majority of these studies has been focused on static Wilson loops, corresponding to non-moving bound states and minimal surfaces with constant boundary conditions over time. In this work we extend several of these studies to rotating Wilson loops and surfaces and we present an analytic explanation of the appropriate renormalization procedures on the energies that need to be applied in this case. These observables are related to minimal surfaces with spinning boundary conditions, corresponding to rotating heavy quark bound states. Such related studies have been initiated in \cite{Peeters:2006iu} where  the rotating meson states were examined in Witten-Sakai-Sugimoto model with a finite radial cut-off and therefore a finite meson mass.

In our current work we initially work in full generality to develop the holographic framework for corresponding surfaces with rotating endpoints. Irrespective of the particular holographic theory the energy of the strings dual to heavy quarks is divergent.  The presence of infinities is due to the infinite distance from the bulk to the boundary of the holographic space and therefore the infinite length of the string worldsheet. The infinite distance from the boundary for the non-rotating case corresponds to the infinite mass of the boundary particles from the field theory perspective.  The renormalization of infinities for the case of moving bound states is significantly more involved compared to the static case and can be performed in different ways generalizing the ideas applied on the static Wilson loop. Namely, we study the holographic renormalization,  which is based on the expansion of the worldsheet energy in the near boundary regime to identify the infinite term and subtract it.  Then we propose the mass renormalization scheme generalizing the static case idea \cite{Maldacena:1998im,Brandhuber:1998bs}. From the energy of the rotating bound state, one removes the infinite thermal mass of the two heavy quark components to remain with the static potential. However, moving and rotating color singlets experience drag in contrast to the color neutral mesons. The drag of the quark is translated in holography with a minimal trailing worldsheet solution that develops itself a black hole horizon. The worldsheet horizon separates the upper segment of the string ending on the boundary of the theory which moves slower than the local velocity of the light, from the lower segment of the string which ends on the horizon of the black hole spacetime and whose local velocity exceeds that of light.  Fluctuations of the string worldsheet horizon in the lower segment are causally disconnected from those on the other side of the horizon, effectively disconnecting this part of the string to its endpoint of the boundary \cite{Fadafan:2008adl,Gubser:2006bz,Casalderrey-Solana:2007ahi}, in any type of holographic theory \cite{Giataganas:2013hwa}. Moreover, the boundary quark experiences an effective temperature that is given by the worldsheet horizon \cite{Giataganas:2013hwa}.  Therefore, the energy subtracted from the total infinite energy of the meson, is the energy of the spiraling string causally connected to the boundary, minus  the energy of the external source that is provided to the color singlet to maintain its motion. We also investigate an approximate renormalization scheme of subtracting a boosted thermal mass, motivated mainly by its simplicity and direct applicability  compared to the accurate mass renormalization scheme. We show however that this approximate renormalization scheme is applicable only for low angular frequencies. 

By fixing the size of the rotating meson $L T$ in a thermal holographic theory of temperature $T$ there are two phenomena that are observed.  There is a maximum at the absolute value of total finite energy with respect to the angular frequencies. Moreover, there is a maximum value of angular frequency beyond which bound states cannot exist. This has been observed also in \cite{Peeters:2006iu} for the rotation of fixed mass mesons. In fact this is a property of the holographic string solutions in thermal backgrounds. Secondly, the comparison of the energy of the rotating meson, renormalized by the infinite thermal mass of the individual quarks minus the energy given by the external source to maintain their motion, leads to a critical angular frequency where the total energy becomes zero. The critical angular frequency is in general lower than the maximum frequency  we have mentioned above, and can be thought as related to the meson melting to its ingredients.

As a side comment to our findings, we note that the holographic renormalization scheme is appropriate to find the maximum frequency beyond which mesons cannot exist. While the mass renormalization scheme requires a cumbersome computation and is appropriate to realise the existence of a critical frequency that the meson melts to its ingredients.

The second part of our work consists on applying the formalism developed above, to a thermal theory of two scales. We choose to study the non-local rotating observable in thermal theories with broken rotational symmetry. The framework for the study of several static observables, as well as their holographic phenomenology in anisotropic theories has been initiated in \cite{Giataganas:2012zy} while there is an extensive followup literature on probes in strongly coupled anisotropic theories and their phase transitions, including for example  \cite{Giataganas:2018uuw,Chernicoff:2012bu,Giataganas:2013hwa,Giataganas:2013zaa,Rajagopal:2015roa, Giataganas:2013lga, Giataganas:2017koz, Giataganas:2018ekx, Arefeva:2020vae,Gursoy:2021efc,Arefeva:2022avn,Golubtsova:2021agl,Bohra:2020qom,Ipp:2020mjc,Iwasaki:2021nrz,DElia:2021yvk,Ipp:2020nfu,DElia:2021tfb}. We firstly solve numerically the supergravity equations of motion to obtain the gravity background. Then we find numerically the string solution with the rotating endpoints of fixed length in the holographic background. The rotating worldsheet becomes parallel to the radial direction in the bulk before reaching the boundary, where it becomes parallel for a second time. We initially work in the static gauge shooting from the turning point of the string. At the point that the string becomes parallel, certain derivatives diverge and we switch to the radial gauge to find the rest of the solution from this point towards the boundary. Using a patching method for the two different segments we find the full string solution. Having obtained the string solutions numerically we can compute their infinite energy. Then we apply the renormalization schemes we have described. The holographic renormalization is relatively straightforward, and provides the maximum angular frequency of the  meson beyond which it cannot exist. We compare the rotating meson states of the same size, in different anisotropies, and we find that the increase of anisotropy reduces the maximum angular frequency $\o$, hinting an easier dissociation of the bound state.

Then we apply the cumbersome mass renormalization scheme. On top of the string worldsheet with the two boundary endpoints that corresponds to the bound state, we  compute the worldsheet that corresponds to the rotating dragging color singlet state \cite{Fadafan:2008adl,Fadafan:2012qu}. For the same fixed radius of the meson, and for a range of angular frequencies we find the solution numerically by shooting from the worldsheet horizon to the boundary, and separately to the black hole horizon. Eventually we patch the two segments to obtain the full solution. We then compute the infinite energy along the string from the boundary to the horizon of the worldsheet, subtracting the energy of the external source that is required to maintain the quark motion. Finally we renormalize the total energy of the meson with the above quantity to obtain the finite energy of the meson dependence on the frequency for different anisotropies. We observe that there is a critical frequency, lower than the maximum frequency mentioned above, at which the renormalized energy becomes zero. This frequency can be interpreted qualitatively as an analogue of the critical frequency that the meson dissociates to its individual quarks. We compare two heavy mesons of the same size at different anisotropies and we find that as the anisotropy increases the critical frequency decreases. Therefore the presence of anisotropy in the theory acts as a catalyst for the melting of rotating mesons with respect to their velocities.

\section{Setup for the Rotation in Holographic Theories}
		
Let us consider a $d+2$-dimensional holographic theory with the generic characteristics of an anisotropic black hole given by
\be \la{metric}
ds^2_{d+2}=g_{tt}(u) dt^2+g_{x_i x_i}(u) dx_i^2+g_{yy}(u) dy^2+g_{uu}(u) du^2~.
\ee
Without loss of generality we consider a $d-$dimensional anisotropic space consisting of a  $(d-1)$-symmetric plane, where the index $i=1,\ldots, d-1$   and an extra dimension $y$ that is responsible for the breaking of the isotropy. The background has a horizon $u_h$ where $g_{tt}(u_h)=0$, and is allowed to have a boundary at $u_b$ where $g_{xx}(u_b)\rightarrow \infty$, which we take without loss of generality to be at $u=0$. The appropriate coordinate system to study the rotation on the isotropic plane is
\be \la{metric2}
ds^2_{d+2}=g_{tt}(u) dt^2+g_{xx}(u) \prt{d\r^2+\r^2 d\phi^2}+\sum_{i=3}^{d-1} g_{x_i x_i}(u) dx_i^2+ g_{yy}(u) dy^2+g_{uu}(u) du^2~,
\ee
where $\phi$ is the cyclic angle. An appropriate parametrization for the rotation of probes and the minimal surface with angular frequency $\o$ in the holographic background is   $t=\tau,~ \rho=\sigma,~ u=u(\sigma),~ \phi=\omega \tau$ which we refer to as a static gauge.  The parametrization is appropriate for the regime of surfaces where the derivative $u'(\s)$ remains finite. The alternative parametrization to which we refer as a radial gauge: $\rho=\rho(\s), ~ u=\s,$ becomes  useful once there are divergences in the static gauge. The use of the two inverse parametrizations is essential for obtaining the full solution of the rotating surfaces since the derivatives $u'(\s)$ diverge in the bulk before reaching the boundary, in contrast to the static probes. We elaborate further on these details later on where we present the surface solutions.  The action in the static gauge reads
\begin{equation}
\prt{2\pi \a'}S_s=  \int d\t d\s \sqrt{-(g_{tt}+g_{xx}\rho^2\omega^2)(g_{xx}+g_{uu}u^{\prime 2})}:=\int d\t d\s \sqrt{D_s} ~, \label{action1}
\end{equation}
where $D_s$ is the square density of the Nambu-Goto action, and let us absorb for the rest of the paper the $\prt{2\pi \a'}$ units in the action for presentation purposes.
The energy and the angular momentum carried by the string are given by the integration of the relevant conjugate momenta along the string as
\be
E_s = \int d\r \frac{1}{\sqrt{D_s}}\prt{-g_{tt}(g_{xx}+g_{uu}u^{\prime 2})}~,\qquad
J_s = \int d\r\frac{1}{\sqrt{D_s}}\omega \rho^2g_{xx}(g_{xx}+g_{uu}u^{\prime 2})~. \label{en1}
\ee
The length of the string is infinite and therefore both of these expressions need a regularization which we discuss later. A no-complex condition has to be imposed on the configuration \eq{action1}, which gives  $\r^2 \o^2\le -g_{tt}/g_{xx}$.
This immediately sets preliminary constraints on the solutions of the equation of motion. For example, the probing string cannot touch the black hole horizon unless it is static or point-like.  

The energy and the angular momentum in the alternative radial gauge can be found in the same way and read
\be \label{en2}
E_r=  \int  du \frac{1}{\sqrt{D_r}}\prt{-g_{tt}(g_{xx}\rho^{\prime 2}+g_{uu})}~,\qquad
J_r = \int du \ \frac{1}{\sqrt{D_r}}\omega\rho^2g_{xx}(g_{xx}\rho^{\prime 2}+g_{uu})~,
\ee
where $D_r$ is the square density of the action in this parametrization
\be
D_r:=\sqrt{-(g_{tt}+g_{xx}\rho^2\omega^2)(g_{xx}\rho^{\prime 2}+g_{uu})}~.
\ee
Notice that the parametrizations presented are consistent for solving the string equations of the rotating probes. The reason is that the meson is color neutral and experiences no dragging phenomena, which are translated to trailing strings.

\subsection{Minimization of Rotating Surfaces}
In this section we elaborate on the anisotropic effects of the rotation in the string probes and describe the strategy to obtain their solutions. The ordinary differential equation of motion for the action \eq{action1} is of second order and reads
\be \la{eom1}
\pp_\s\prt{\ff{1}{\sqrt{D_s}}u'g_{uu}\prt{g_{tt}+g_{xx}\r^2\o^2}}-\ff{1}{2\sqrt{D_s}}\pp_u\prt{(g_{tt}+g_{xx}\rho^2\omega^2)(g_{xx}+g_{uu}u^{\prime 2})}\ = 0~.
\ee
The boundary conditions correspond to the two endpoints of the string at the boundary of the holographic space, defining the radius $L$ of the rotating meson $u(\pm L)=0$ and requiring the string to have a turning point in the bulk such that $u'(\s_0)=0$, which by symmetry for the type of our solutions should be located at $\s_0=0$.

There are few immediate remarks on the string solutions. The reasons that the parametrizations we use are enough and consistent to describe the rotation of the string is solely due to the fact that the dual mesons are color neutral. Therefore we expect that they will not experience any drag effects as happens with the color singlets and there is no need to apply a constant force on the boundary to maintain the rotation of the meson. This is in contrast to the rotation of color singlets where an external force is necessary to maintain their motion, and the momentum flux from the boundary to the bulk along the string results to the generation of a black hole horizon on the induced string metric  \cite{Gubser:2006bz}, which depends on the anisotropy and leads to universal phenomena \cite{Giataganas:2012zy,Giataganas:2013hwa,Giataganas:2013zaa}. Therefore the rotating string profile corresponding to the meson can be described as a rigid one with no trail. Nevertheless, we find that the angular frequency does generate deformation effects on the string profile due to rotation which depend on the magnitude of the angular velocity \cite{Peeters:2006iu,Chu:2016pea}, in contrast to the static string in thermal theories.
		
This string deformation due to finite angular momentum complicates the strategy of obtaining the numerical string solution. The string always reaches the boundary orthogonally satisfying $u'(\r) \rightarrow \infty$ which can be also understood by the fact that the string extends infinitely from the bulk to the boundary. For zero angular momentum the numerical treatment of equation \eq{eom1} is enough to obtain the full string solution, with its boundary endpoints lying at  $\r=\pm L$, where along the string the "radius"  satisfies $\r^2 \le L^2$. For a finite angular momentum there is a string deformation that $\r \ge L$ in the bulk, while on the boundary $\r=\pm L$. This leads to a new numerical divergence, since there exists a saddle point in the bulk at $u(\s_1)$, that the deformation maximizes such that $u'(\s_1)\rightarrow \infty$. Therefore, we obtain numerically the string solution from the turning point of the string $u(\s_0=0)$ to $u(\s_1)$ by solving \eq{eom1}, where we change parametrization to the radial gauge and we solve the inverse equation which reads
\be \la{eom2}
\pp_\s\left[\frac{1}{\sqrt{D_r}}\rho^{\prime}g_{xx}(g_{tt}+g_{xx}\rho^2\omega^2)\right]-\ff{1}{2\sqrt{D_r}}\partial_{\rho}\prt{(g_{tt}+g_{xx}\rho^2\omega^2)(g_{xx}\rho^{\prime 2}+g_{uu})}=0~.
\ee
The above equation is solved from $u(\s_1)$ to the boundary, using the initial conditions obtained from the static gauge equation \eq{eom1} at $u(\s_1)$.  Following this strategy we obtain the full string solutions in holographic theories as we apply it in later sections. It may be instructive for the reader to have a first look at the Figures \ref{figure:a0} and  \ref{figure:a1}, in order to visualize the shape deformations we describe in this section and the need of the two gauges. The different colors on the string in Figures represent the two different gauges we have used to obtain the relevant solution.

The energy and the momentum carried by the string is given by the integrals \eq{en1}  for the range $\r=0$ to $\r=\s_1$ for the solution \eq{eom1}, and the integrals in  \eq{en2} from $u(\r=\s_1)$ to $u(\r=L)$ for the solution \eq{eom2}. The total energy and momentum is given by their sum, while to take into account the whole symmetric string with the integration boundaries we mention we need to multiply by two:
\be \la{etot}
E_{bound}= 2\prt{ E_s+E_r}~,\qquad J_{bound}= 2 \prt{ J_s+J_r}~.
\ee
The length of the string is infinite,  therefore the above expressions are infinite  and require renormalization.

\section{Renormalization of Infinities for Rotating Probes} \la{section:renor}

The energy of the holographic strings dual to heavy quarks is divergent.  The presence of infinities lies on the infinite distance from the bulk to the boundary in the holographic space and therefore the infinite length of the string itself. The infinite string distance from the boundary for static quarks corresponds to the infinite mass of the boundary particles, from the field theory perspective.  The renormalization of infinities can be performed in different ways we suggest below.

A straightforward approach for static bound states consists of subtracting the bare mass of the quark. In particular, for static mesons  the corresponding counterterm that is added in the energy to cancel the UV divergence, corresponds to the static solution of the straight string and it is the dominant contribution of the infinite mass of the static quark
\be\la{ren0}
2 \pi \a' S_m=2 \cT\int_{u_k}^{u_b}du \sqrt{-g_{tt} g_{uu}}:=S_{UV}(u_b)-S_{IR}(u_k)~,
\ee
where $u_b$ is the boundary and $u_k$ is the deepest point in the bulk of the spacetime, for example the position of the horizon of the black hole in case that it is present. $\cT$ is the time scale of the integration and does not play any role in the computation. The counterterm $S_{UV}$ depends only on the intrinsic variables of the theory and on the cut-off $u_b$. It is equal to the upper bound contribution of the above integral. The whole thermal mass $S_m$ depends on the state of the theory. The renormalization scheme for such probes has been proposed in \cite{Maldacena:1998im,Brandhuber:1998bs} and is applicable in anisotropic theories \cite{Giataganas:2012zy} we examine below. This is a widely used and straightforward renormalization scheme for the static probes, while for the moving ones there are certain complications.

The extra complication comes from the fact that color singlets are charged and experience drag phenomena while moving in thermal field theories. As a result the quark corresponding to the string solution with a single endpoint needs the application of a constant external force to keep it moving, resulting to a momentum flow from the boundary to the bulk generating a trailing string solution with an intrinsic black hole in the worldsheet. Such solutions have been examined in depth  in isotropic \cite{Gubser:2006bz,Casalderrey-Solana:2007ahi,Gursoy:2010aa}  and generic holographic theories  \cite{Giataganas:2012zy,Giataganas:2013hwa,Giataganas:2013zaa,Fadafan:2012qu}. Integrating the energy over the  trailing rotating string solutions will result to the cancellation of the divergences of the energy \eq{etot}, however will add the contribution of the dragging of the color singlet, which does not have a counterpart in the color neutral meson.

Note that the divergences for rotating probes appear in the angular momentum as well. The reason is the same as in the energy, and the schemes to renormalize the infinities are parallel to the ones we apply for the energy. Below we describe the most natural renormalization schemes for the energy and momentum, which can be thought eventually as corresponding to different type of observables with a center the quarkonium energy.

\subsection{Holographic Renormalization} \la{section:holren}
Following our previous discussion, an appropriate scheme to cancel the divergences in the energy is to take into account the rotation by considering the velocity dependent counterterm analogue of the $S_{UV}$, which we call $S_{UV,hol}$ such that the finite energy is
\be \la{ren1}
E_{tot}= E_{bound}-S_{UV,hol} ~,\qquad S_{UV,hol}=f(\o L) S_{UV}(u_b)~,
\ee
where $E_{bound}$ is given by \eq{etot} and $f(\o L)$ is determined by the characteristics of the probe solely on the boundary and depends on the velocity. The idea behind this scheme is to expand the energy in the near boundary regime, identify the infinite term and subtract it, as we do for the holographic theories counterterms  \cite{deHaro:2000vlm}.

The same type of divergences appear in the angular momentum as well. The reason is the same as in the energy and the schemes to renormalize the infinities are parallel to the ones we apply for the energy to get
\be \la{jren1}
J_{tot}= J_{bound}-J_{UV,hol} ~,\qquad J_{UV,hol}=h(\o L) S_{UV}(u_b)~,
\ee
where $h(\o L)$ is determined by the characteristics of the probe on the boundary  and depends on the velocity. To determine the form of these terms we expand \eq{en2} asymptotically on the boundary $\r=L$ and by using the fact that the string approaches it perpendicularly we get
\be \la{ren_hol0}
S_{UV,hol}\simeq \int^{u_b} du\sqrt{\ff{-g_{tt} g_{uu}}{1+\ff{g_{xx}}{g_{tt}}L^2 \o^2}}~,\qquad  ~J_{UV,hol}\simeq L^2 \o \int^{u_b} du\sqrt{\ff{g_{uu}}{-g_{tt}}}\ff{g_{xx}}{\sqrt{1+\ff{g_{xx}}{g_{tt}}L^2 \o^2}}~.
\ee
The above expressions take into account the rotation effects on the energy and momentum, and the divergences appear with the right factor in order to obtain a finite expression in \eq{etot}. At the limit $\o=0$  one recovers  the static counterterm $S_{UV}(u_b)$ that is included in the equation \eq{ren0}.
Here we will focus on non-trivial (anisotropic) RG flows that have a conformal UV fixed point. For theories with a conformal UV fixed point  the near boundary  expansion \eq{ren_hol0} gives
\be \la{ren_hol1}
S_{UV,hol}\simeq \ff{1}{\sqrt{1-L^2 \o^2}} S_{UV}(u_b)~,\qquad  ~J_{UV,hol}\simeq \ff{ L^2 \o }{\sqrt{1-L^2 \o^2}} S_{UV}(u_b)~, 
\ee
where the $S_{UV}(u_b)$ is the same of the \eq{ren0}.
In summary, for the holographic renormalization scheme the finite energy and momentum for the rotating bound state of radius $L$ and angular frequency $\o$, is given by the equations \eq{ren1}, \eq{jren1} where the counterterms in the expressions are given by the equation \eq{ren_hol1}.

\subsection{Approximate Boosted Mass Renormalization}  \la{section:bmass}

The above expressions \eq{ren_hol1} would match the expected relativistic energy and momentum, if instead of $S_{UV}(u_b)$ we had the $S_m$, the thermal mass of the quarks, since the Lorentz factor appears naturally.  In fact this motivates the computation of the following finite quantity as an approximate scheme to observe the phase transitions of rotating mesons to their free ingredients
\be \la{ren_mass}
E=\prt{E_{bound}-S_{UV,hol}}-\prt{ \ff{1}{\sqrt{1-L^2 \o^2}} S_{m}-S_{UV,hol}}= E_{bound}- \ff{1}{\sqrt{1-L^2 \o^2}} S_{m}~,
\ee
where $S_{m}$ is the thermal mass of the two static quarks, and in this expression is approximated as given by \eq{ren0}. A similar renormalization formula applies to the angular momentum. Although this is close to be the analogue of the static mass subtraction scheme, there is a significant complication we have neglected in writing down the equation \eq{ren_mass}. A moving straight string extending from the boundary to the horizon of the black hole, corresponding to the $S_{m}$ in \eq{ren_mass},  is not a solution in black hole environments, due to the fact that the color singlet is sensitive  to dragging phenomena and therefore energy loss occurs.  The expression \eq{ren_mass} does not capture such phenomena and is not accurate. Nevertheless, it can be thought as serving as a good and straightforward approximation of the exact color singlet renormalization moving in the thermal background at least for low velocities and it is worthy to be briefly discussed. We will compute the range of the validity of this approximate scheme in the following sections. An alternative motivated discussion on this type of term and its approximate nature interpreting it as a mass term in a different setup appears in \cite{Herzog:2006gh}.

\subsection{Rotating Color Singlet Mass Subtraction Renormalization} \la{section:spiral}

The discussion of the previous sections motivates the most natural subtraction scheme to observe the phase transitions from a rotating meson bound state to its color singlets. In this scheme, the energy and momentum of a meson of size $2L$ and angular frequency $\o$, is compared with the energy and momentum of its two ingredients. That is two free quarks moving along a circle of radius $L$ with the same angular frequency. The color singlet corresponds to a rotating string with a single boundary endpoint. The string worldsheet due to the drag phenomena develops a non-trivial trailing profile. Such rotating strings have been studied in isotropic and anisotropic theories in \cite{Fadafan:2008adl,Fadafan:2012qu} and can be thought as generalization of linear string motion \cite{Gubser:2006bz,Giataganas:2013hwa,Giataganas:2013zaa} \footnote{Another  direction using our setup would be to study the analog of the Wilsonian renormalization of rotating observables to derive the holographic effective string action \cite{Gutiez:2020sxg}.}. The trailing worldsheet develops a horizon $u_{ws}$ which depends on the velocity of the quark at the boundary and the other scales present in the theory. Thus one may consider as an interesting physical quantity the following
\be \la{ren_bend1}
E=\prt{E_{bound}-S_{UV,hol}}-\prt{E_{singlet} -S_{UV,hol}}= E_{bound}-E_{singlet} ~.
\ee
Finding $E_{singlet} $ is a non-trivial exercise. We first need to find the numerical solution of the spiraling string with a single endpoint at the boundary.  The string is parametrized in the radial gauge with $\phi(u) = \o \t +\th(u)$. The function $\th(u)$ parametrizes the spiraling of the trailing string and it is necessary to caption the right worldsheet dynamics.  
The action for the spiraling string is given by
\be\la{rot}
S_{singlet} =\int du \sqrt{-\prt{g_{tt}+g_{xx}\r^2 \o^2}\prt{g_{uu}+g_{xx}\r'^2}-g_{tt} g_{xx}\r^2\th'^2 }:= \int du\sqrt{D_{singlet}}~,
\ee
where $D_{singlet}$ is the square density of the Nambu-Goto action for the spiraling string, and we have absorbed the $(2\pi \a'/\cT)$ in the action following our previous conventions. The equation of motion for $\th(u)$ reads
\be\la{thprime1}
\th'^2=-\P^2\ff{\prt{g_{tt}+g_{xx}\r^2\om^2}\prt{g_{uu}+g_{xx}\r'^2}}{g_{tt}g_{xx}\r^2\prt{g_{tt}g_{xx}\r^2+\P^2}}~,
\ee
where $\Pi$ is a constant of motion, and can be thought as the rate of energy loss of the string. It is the energy required to keep the quark moving with a constant angular velocity. Since it is constant it can be computed at any point of the trajectory and the most convenient point is at the string worldsheet horizon $u_{ws}$ .  
We obtain $u_{ws}$  by looking at the sign change point of the above ratio \eq{thprime1} which leads to the following algebraic system
\be\la{ucc}
g_{tt}(u_{ws})=\P\om~, \qq g_{xx}(u_{ws})\r(u_{ws})^2=-\ff{\P}{\om}~,
\ee
which can be solved numerically.  $u_{ws}$ depends on $\Pi, \o$ and the rest of the scales of the theory. It is the horizon of the worldsheet that separates the segment of the string that ends on the boundary of the theory which moves slower than the local velocity of the light, from the segment of the string which ends on the horizon of the black hole spacetime and whose local velocity exceeds that of light. This is analogous to the linear quark motion. The equation for the radius profile of the string
\be\la{rps}
\ff{\r g_{xx}\prt{\o^2\prt{g_{uu}+g_{xx}\r'^2}+g_{tt}\th'^2}}{\sqrt{D_{singlet}}}-\pp_\s\prtt{\ff{g_{xx} \r'\prt{g_{tt}+g_{xx}\r^2 \o^2}}{\sqrt{D_{singlet}}}}=0~,
\ee
will be solved numerically together with equation \eq{thprime1}. We again use a type of patching method to obtain the full string solution since at $u_{ws}$ there is a singularity. We  solve perturbatively \eq{rps} at $u_{ws}$ to determine the first derivative $\r'$ in this regime and therefore the initial conditions for the differential equation. Then we shoot from this point towards the boundary of the background and then towards the horizon of the black hole to obtain the two segments of the string. The energy and the angular momentum carried by the string are given by the integration of the relevant conjugate momenta along the string as
\ba\la{enermom}
E_{singlet}=2\int_{u_{ws}}^{u_b}du \ff{  -g_{tt}\prt{g_{uu}+g_{xx}\r'^2+ g_{xx}\r^2 \th'^2}}{\sqrt{D_{singlet}}}~,\qquad J_{singlet}=2\int_{u_{ws}}^{u_b}du \ff{ \o g_{xx}\r^2\prt{g_{uu}+g_{xx}\r'^2}}{\sqrt{D_{singlet}}}~.
\ea
The dependence of $\th'$ can be eliminated by the use of the equation \eq{thprime1}. Knowing the profile of the string we can integrate along the string to find its infinite energy and momentum so that to renormalize the observable \eq{ren_bend1}. This quantity can be thought as the closest analogue to the mass renormalization scheme for the static heavy quark bound states\cite{Maldacena:1998im}.

Nevertheless note that in order for the two color singlets to have the same boundary conditions $(L,\o)$ with the bound state meson, a continuous application of a constant force is needed in order to maintain its rotation counterbalancing the energy loss phenomena of the quark in the thermal environment. One may also subtract from \eq{ren_bend1} the energy deposited in the medium for a fixed unit of time $\cT$. This is equal to the energy of the external source responsible for maintaining the motion of the quark with a constant speed. This energy loss for a period of time $\cT$ (which is the same period we have computed and normalized the other energies in the manuscript) is given by
\be \la{external}
E_{external}=- 2 g_{tt}(u_{ws})\ff{1}{2\pi \a'}~,
\ee
while we will again  absorb the units of $2 \pi\a'$ to be consistent in our notation and the factor of two is due to the two quarks. It is instructive to subtract the energy of the external source to compute the quantity
\be \la{ren_bend12}
E=\prt{E_{bound}-S_{UV,hol}}-\prt{E_{singlet} -S_{UV,hol}- E_{external}}= E_{bound}-\prt{E_{singlet}- E_{external}} ~.
\ee
This is the way we choose to renormalize the energy. In summary the mass renormalization scheme is obtained by equation \eq{ren_bend12}, where $E_{external}$ is given by  \eq{external}, and $E_{singlet}$ is given by the equation \eq{enermom} integrated along the string solution of the differential equation \eq{rps} for the string segment above the worldsheet horizon which is given by the algebraic solution of equations \eq{ucc}.

\section{Brief Implementation of the Analytic Relations}

In this section we warm up by presenting a brief analysis on an analytic perturbed background without specifying explicitly the theory. The scale of the perturbation is $a$, which can be thought as related to a source that generates an anisotropy. The metric for such a theory takes perturbatively the form
\bea\nn
&&g_{tt}(u)= -g_{xx0}(u) f(u)-a^2g_{tt2}(u)~,\quad
g_{uu}(u) = \frac{g_{xx0}(u)}{f(u)}+a^2 g_{uu2}(u)~,\quad
g_{x_1 x_1}(u) = g_{xx0}(u) ~,\\
&&g_{x_2 x_2}(u,\r)= g_{xx0}(u)\r^2~,\quad
g_{yy}(u) = g_{xx0}(u)+a^2g_{yy2}(u)~,
\eea
where $f(u)$ is the blackening factor of the generic black hole with temperature $T$. Then $a\ll T$ and $a\ll (\mbox{any other scale in the theory}$) can be considered as the low anisotropy limit where the leading terms in the expansion correspond to an isotropic thermal theory, and
\be
u(\s)=u_0(\s)+a^2 u_2(\s)~,\quad \r(\s)=\r_0(\s)+a^2 \r_2(\s)~,
\ee
are the solutions $u(\r)$ and $\r(u)$ that correspond to the static and radial gauge respectively. The energy and momentum is modified from the isotropic theory in the static gauge as
\begin{eqnarray}\nn
	E_s&= &\int d\r \bigg[\frac{g_{xx0}\sqrt{f\prt{f+u_0'{}^2}}}{\sqrt{f-\rho^2\omega^2}}+
	a^2 \ff{f \prt{f+u'_0{}^2}}{2 \prt{f  \prt{f+u'_0{}^2}
		\prt{f-\rho ^2 \o^2}}{}^{3/2}}\cdot \left(\prt{f+u'_0{}^2} \left(f-2\rho^2 \o^2\right) g_{tt2}+\right.  \\&&\hspace{-1.2cm}
\left.u_2 f' g_{xx0}  \left(f \left(f-2 \rho ^2 \o^2\right)-\rho ^2 \o^2 u'_0{}^2\right)+f \left(f-\rho ^2 \o^2\right) \left(u'_0 \left(f u'_0 g_{uu2}+2 g_{xx0} u'_2\right)+2 u_2 \prt{f+u'_0{}^2} g'_{xx0}\right)\right)\bigg]~,\nn\\ \nn
J_s &=&\int d\r \bigg[ \frac{\omega\rho^2
	g_{xx0}\sqrt{f+u'_0{}^2}}{\sqrt{f \prt{f-\rho^2\omega^2}}}-a^2\ff{\rho ^2 \o 
	\left(f+u'_0{}^2\right)}{2 \prt{f  \prt{f+u'_0{}^2}
		\prt{f-\rho ^2 \o^2}}{}^{3/2}}\cdot \left(f \left(f+u'_0{}^2\right) g_{tt2}+\right.\\
&&\hspace{-1cm}\left.u_2 f' g_{xx0} \left(f^2+u'_0{}^2 \left(2 f-\rho ^2 \o^2\right)\right)-f \left(f-\rho ^2 \o^2\right) \left(u'_0 \left(f u'_0 g_{uu2}+2 g_{xx0} u'_2\right)+2 u_2 \left(f+u'_0{}^2\right) g'_{xx0}\right)\right)\bigg] ~,\nonumber
\end{eqnarray}
and in the radial one as
\begin{eqnarray}\nn
	E _r
	&=&\int du \bigg[\frac{g_{xx0}\sqrt{1+f\r'_0{}^2}}{\sqrt{1-\frac{\rho_0^2\omega^2}{f}}}+a^2\ff{f  \left(f \rho'_0{}^2+1\right)}{2 \left(f  \left(f \rho'_0{}^2+1\right) \left(f-\rho _0^2 \o^2\right)\right){}^{3/2}}\cdot \left(f^2 \left(f-\rho _0^2 \o^2\right) g_{uu2}\right.\\&&+ \left.\left(f \rho'_0{}^2+1\right)\left(f-2 \rho _0^2 \o^2\right) g_{\text{tt2}}+2 f g_{xx0} \left(\rho _0 \rho _2 \o^2 \left(f \rho'_0{}^2+1\right)+f \rho '_0 \rho '_2 \left(f-\rho _0^2 \o^2\right)\right)\right)\bigg]~, \nonumber \\\nn
	J_r &=&\int du\bigg[ \frac{\omega\rho_0^2 g_{xx0}\sqrt{1+f  \r'_0{}^2}}{f\sqrt{1-\frac{\rho_0^2 \omega^2}{f}}}-a^2\ff{f \rho _0 \o  \left(f \rho'_0{}^2+1\right)}{2 \left(f \left(f \rho'_0{}^2+1\right) \left(f-\rho _0^2 \o^2\right)\right){}^{3/2}}\cdot \left(\rho _0\left(f \left(\rho _0^2 \o^2-f\right) g_{uu2}\right.\right.\\&&+\left.\left.\left(f \rho'_0{}^2+1\right) g_{\text{tt2}}\right)+2 g_{xx0} \left(\rho _2 \left(f \rho'_0{}^2+1\right) \left(\rho_0^2 \o^2-2 f\right)+f \rho _0 \rho'_0 \rho'_2 \left(\rho_0^2 \o^2-f\right)\right)\right) \bigg]~,\nonumber
\end{eqnarray}
where the anisotropic contribution enters in the subleading terms. Once the background is known the above expressions in certain circumstances can be conclusive on the qualitative effect of the anisotropy of the rotation of the probes. For theories where the sign of the perturbation integrand is definite along the whole RG flow, one should be able to extract a conclusive qualitative answer on the effect of the anisotropy  without evaluating the integrals. In the following sections we apply our framework on certain theories numerically presenting a precise analysis.

\section{Applications on the Axion Deformed Anisotropic Theories} \la{section:aniso}
		
To apply our formalism let us consider a theory with multiple scales so that the observations on the rotating probes are more involved.  Let us consider backgrounds that are derived by the IIB supergravity action in the string frame
\be \la{eqlen1}
S=\ff{1}{2 \k^{2}_{10}}\int_\cM d^{10} x\sqrt{-g} \prtt{e^{-2\phi}\prt{R+4\pp_M \phi \pp^M \phi}-\ff{1}{2} F_1^2-\ff{1}{4\cdot 5!}F_5^2}~,
\ee
where the index $M=0,\ldots,9$, $F_1=d\chi$ is the axion field strength and $\phi$ denotes the dilaton. The action is known to have solutions of the form \cite{Azeyanagi:2009pr,Mateos:2011ix}
\bea\la{metrica1}
&&ds^2=\ff{1}{u^2}\prt{-\cF(u) \cB(u) dt^2+dx_1^2+dx_2^2+\cH(u) dx_3^2+\ff{du^2}{\cF(u)}}+\cZ(u) d\Omega_{S^5}^2 \nn \\
&&\chi=a x_3~, \quad \phi=\phi(u)~,
\eea
with an asymptotic AdS boundary. The anisotropic RG flows can be found numerically, while  analytical solutions are possible in the limit of low $a/T$. At the order of $a^2$ the solution reads
\ba\nn
&&\cF(u) = f(u)+a^2\cF_2(u)~,\quad
\cB(u) = 1+a^2\cB_2(u)~, \quad
\cH(u) = e^{-\f(u)}~, \quad \cZ(u)=e^{\ff{\f(u)}{2}}~,\\ \nn
&& \f(u)=a^2\f_2(u)~,\quad f(u):=1-\ff{u^4}{u_h^4}  ~,\quad u_h=\ff{1}{\pi T}+\ff{a^2}{T^2}\ff{5\log 2-2}{48\pi^3T}~,
\ea
where the functions are
\ba
\cF_2(u) &=& \ff{1}{24u_h^2}\left[8u^2(u_h^2-u^2)-10u^4\log 2+(3u_h^4+7u^4)\log \prt{1+\ff{u^2}{u_h^2}}\right]~, \nonumber \\\nn
\cB_2(u) &=& -\ff{u_h^2}{24}\left[\ff{10u^2}{u_h^2+u^2}+\log \prt{1+\ff{u^2}{u_h^2}}\right]~, \qquad
\f_2(u) = -\ff{u_h^2}{4}\log \prt{1+\ff{u^2}{u_h^2}}~.
\ea
In our study we focus on the numerical solutions of the form \eq{metrica1} that are exact and valid for any value of anisotropy.  We solve numerically the equations of motion derived by the action \eq{eqlen1} for the background \eq{metrica1} for large anisotropies beyond the perturbative regime,   to obtain the functions $\cF(u), ~\cB(u),~ \cH(u),~ \cZ(u),$ and $\phi(u)$ with UV conformal boundary conditions and anisotropic Lifshitz IR, as in \cite{Mateos:2011ix}. The equation of motion for the dilaton obtained by the action  \eq{eqlen1}, can be expressed after some manipulations as a third order differential equation that is independent of the other functions. It is solved perturbatively in the near horizon regime with an expansion of the form $\phi=\phi_h+ \phi_n (u-u_h)^n$.  Once the dilaton is solved we proceed to solve the rest of equations. We use the perturbative dilaton solution as initial data to start shooting  from the horizon of the black hole to numerically obtain the rest of the background functions. By completing all the integrations we obtain smooth numerical data as a function of the holographic direction describing  the complete RG flow for the desirable choices of $a/T$. In our setup the dual theory has a conformal UV fixed point and an anisotropic Lifshitz IR fixed point with constant anisotropic exponent $z=3/2$. It turns out that the value of the exponent is fixed by the form of the axion-dilaton coupling in the supergravity action. Anisotropic theories with non-trivial RG flows and arbitrary exponents,  describing confining-deconfining phase transitions and their dependence on the anisotropy have been obtained in \cite{Giataganas:2017koz}  and anisotropic theories with trivial RG flows in \cite{Giataganas:2018ekxv,Inkof:2019gmh}.

\subsection{Properties of Rotating Minimal Surfaces} \la{aniso_surfaces}

In this section we elaborate on our strategy to find the spinning strings solutions in the anisotropic theory numerically. We briefly describe the general methodology to solve the  equations of motion for such strings in any holographic background.  We like to fix the boundary length of the string while varying the other parameters of the theory. In this way  we will be able to obtain valid conclusions on the effects of angular frequency and anisotropy on the bound states of fixed size.  The turning point of the string  which is parametrized with the static gauge, is related to the boundary conditions of the string which are described by the radial gauge. 
For a range of angular velocities, we determine the set of the turning points $u_\star(\o,LT)$, where $LT$ is kept fixed for certain values of anisotropy. 
Note that $u_\star$ eventually will be the part of the initial conditions in the static gauge numerics, while $L$ is relevant to the part of the string parametrized by the radial gauge.

The full string solution is obtained by a patching method of the two segments of the string. Firstly we integrate numerically the differential equation  \eq{eom1} in the static gauge starting from the turning point $u_\star(\o,LT)$ to obtain a function $u(\r)$. For $\o\neq 0$ the derivative $u'(\r)$ blows up in the bulk before the string reaches the boundary, since there exists a point in the bulk that $\r>L$. This deformation is caused by the angular momentum. At this point we switch to the radial gauge which is more convenient numerically and we shoot from there to the boundary solving the equation \eq{eom2} to obtain $\r(u)$. The patching of the two segments in the two different gauges gives the full string solution which reaches the boundary at a fixed value $LT$. Similar numerical approaches have been used to obtain rotating strings in dS slicing of AdS \cite{Chu:2016pea}.

\begin{figure}[!t]
	\begin{minipage}[ht]{0.5\textwidth}
		\begin{flushleft}
			\centerline{\includegraphics[width=70mm]{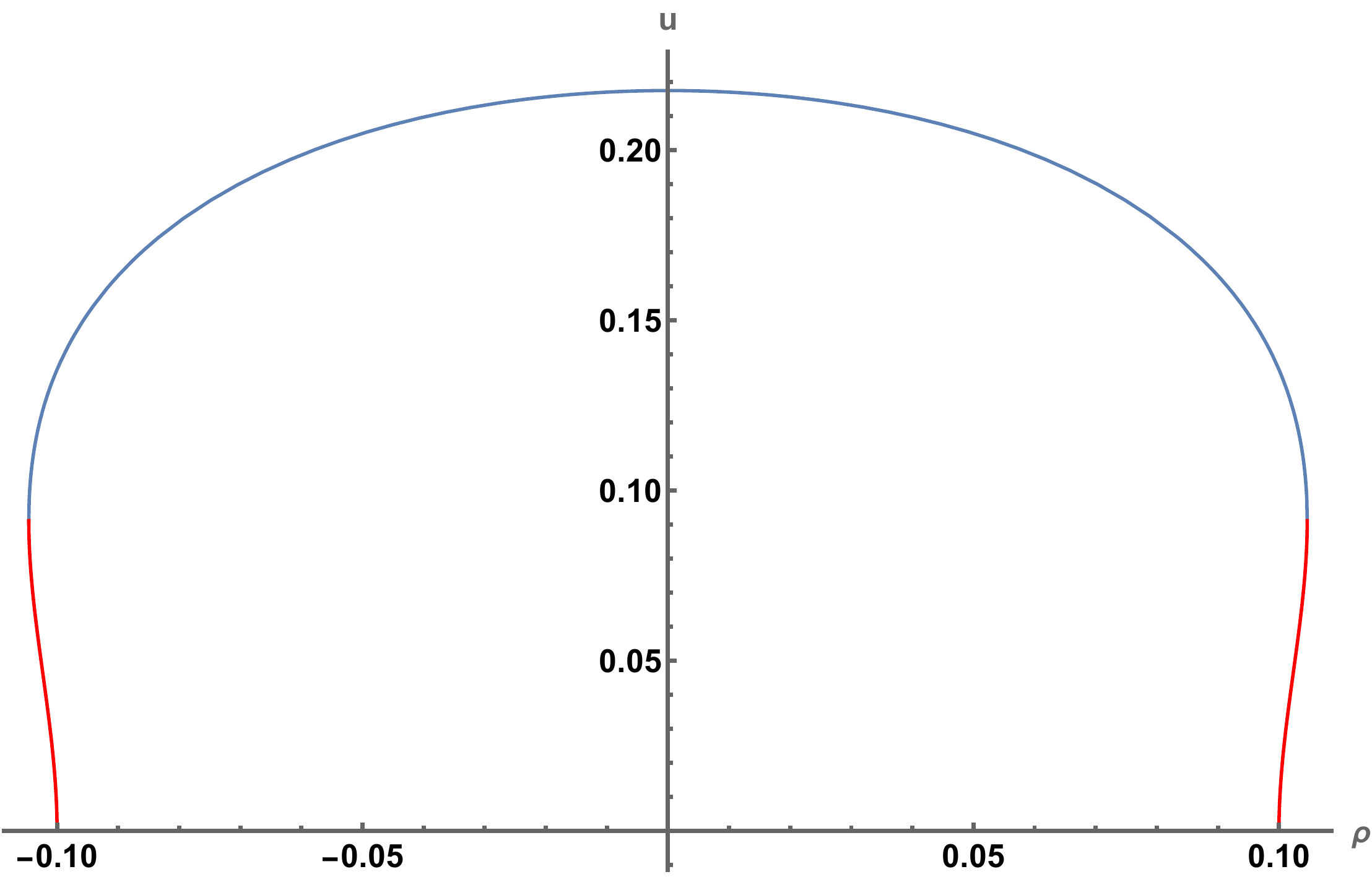}}
			\caption{\small{A string solution of the differential  equations \eq{eom1} and  \eq{eom2} corresponding to a rotating bound state. We shoot from the turning point of the string towards the boundary in the static gauge to solve the equation \eq{eom1}. The string develops a local saddle point and the derivative $u'(\r) $ blows up. At this point we switch to the more convenient radial gauge.  We shoot from this point with the right initial conditions towards the boundary solving  \eq{eom2}. This full string is obtained by patching the two solutions. The solution of the radial gauge is depicted with red color and in the static with blue. The turning point has been found by the methods described in the main text such that the boundary length $L T=.1$, while in this plot the temperature, anisotropy, and spin are kept constant. }}
			\label{figure:a0}
		\end{flushleft}
	\end{minipage}
	\hspace{0.3cm}
	\begin{minipage}[ht]{0.5\textwidth}
		\begin{flushleft}
			\centerline{\includegraphics[width=77mm ]{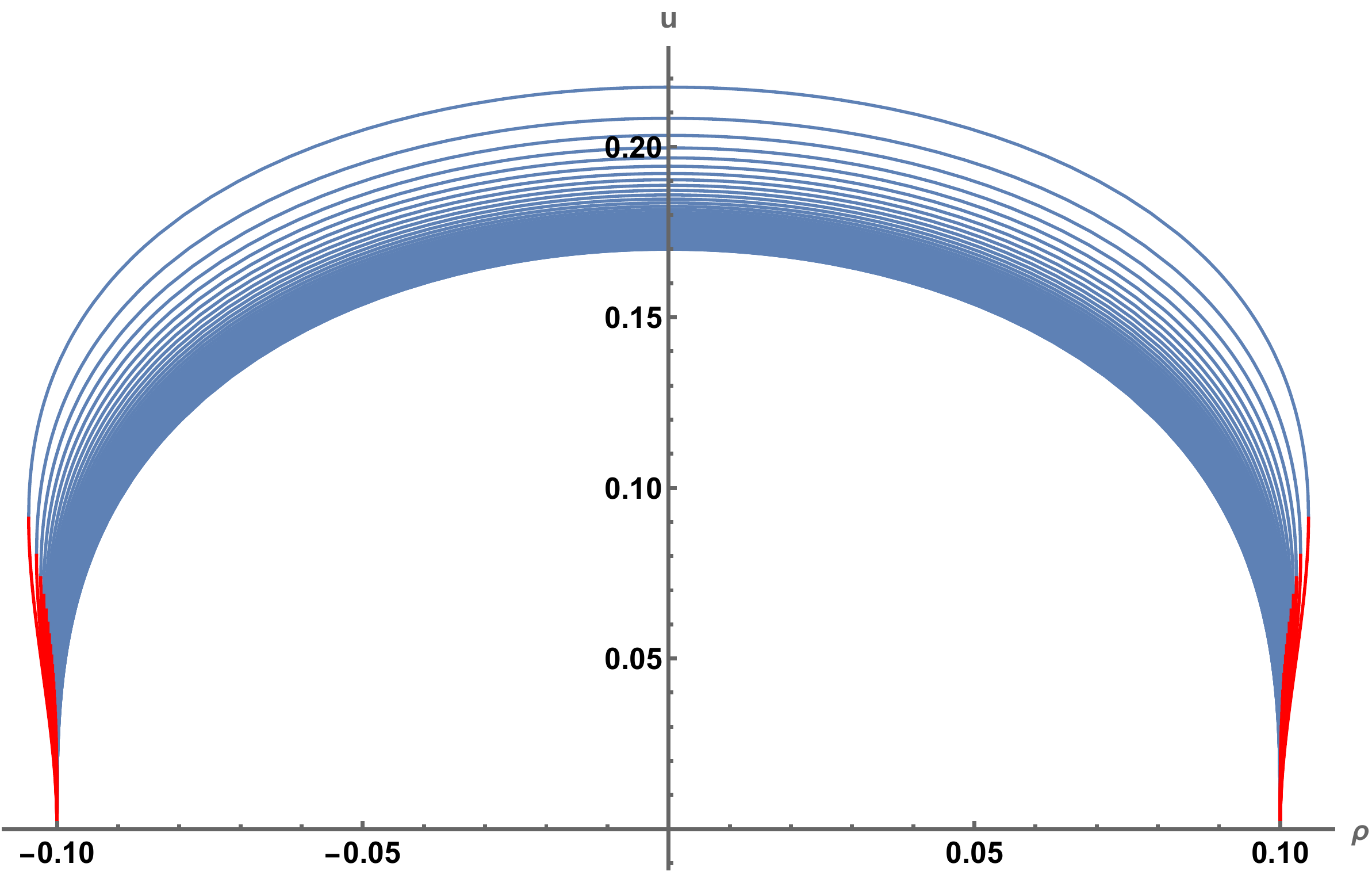}}
			\caption{\small{The spinning strings at anisotropy $a/T=0.5$ and fixed radius $LT=.1$. The solutions are for angular velocities ranging from almost static strings to $\o\simeq4.9$, which is approximately the faster spin that is allowed for the chosen boundary data, such that a solution of the equations \eq{eom1} and  \eq{eom2} exists. The slowest rotating string is the one that remains closer to the boundary, while the fastest one reaches deeper in the bulk. Increase of the velocity deforms stronger the string and moves the point $u'(\r)\rightarrow \infty$ deeper in the bulk as it is depicted by the extension of the regions parametrized with the radial gauge plotted with red color.} }
			\label{figure:a1}\vspace{1.5cm}
		\end{flushleft}
	\end{minipage}
\end{figure}

The results of the numerical integration described above are presented in Figures \ref{figure:a0} and  \ref{figure:a1}. The string segments in the two different gauges are presented with different coloring. In the zero velocity limit one may work only in the static or radial gauge since the string deformation is absent, i.e. $\r\le L$ and the  divergence in the derivative occurs at the boundary. As the angular velocity increases the deformation effects are stronger. This can be understood in the plots by noticing the increase of the segment of the string parametrized by the radial gauge and also the obvious deformation of the shape of the string.  Moreover, the increase of the velocity forces the string solution to extend deeper in the bulk in order to keep fixed its boundary length.

Since we have now the tools to obtain the full string solution for any anisotropy we can proceed to compute the  energy and the momentum along the string.
		
\subsection{Energy  of Moving Probes in  Anisotropic Theories} \la{section:lyap}

In this section we compute the energy of the rotating probes in anisotropic theories in the different renormalization schemes, which we also compare to each other.

\subsubsection{The Numerical Methods}\la{section:num}

Let us make a description of the numerics applied on the anisotropic theory. We first determine the holographic theory numerically. We choose to work with the physical choice of fixing the scales of our theory  $a/T$ and produce the background for several values of the ratio.  We solve numerically the equations of motion derived by the action \eq{eqlen1} for the background \eq{metrica1} to obtain the functions $\cF(u), ~\cB(u),~ \cH(u),~ \cZ(u),~ \phi(u)$. The method has been described already in section \ref{section:aniso} below the metric ansatz.

Once we have obtained the background theory we proceed to solve the rotating string worldsheet equation with the two endpoints on the boundary following the methodology described in the previous subsection. We do it as following, for each pair $(\o, a/T)$ we find the initial data for the string differential equations and more particularly its turning point $u_\star$,  such that $L T= \mbox{constant}$. In this sense we obtain the set of values $u_\star(\o,a/T)$ such that $L T= \mbox{constant}$. Our primary target is to compute the energies of the different renormalization schemes for different anisotropies while keeping the size of the meson fixed. 

Using the obtained $u_\star(\o,a/T)$ data we derive the string solutions for the range of angular frequencies $\o$ and the different values $a/T$, following the methods we have described in subsection \ref{aniso_surfaces}, and patching each segment of the string solution parametrized by the radial and static gauge.  At this point we have obtained the desirable full string solutions. Then we need to integrate the energy and the momentum expressions along  the string in the static gauge and radial gauge respectively using the expressions  \eq{en1} and \eq{en2}.  The string segment in the radial gauge is the one that reaches the boundary and has to be renormalized using the suggested renormalization schemes we describe above. In practice, a numerical regulator is used very close to the boundary which  does not affect the final result since the regulator is canceled for the observables we compute. 

The holographic renormalization and the approximate mass subtraction scheme can then be applied in a relatively straightforward way, by applying the formulas of the sections \ref{section:holren} and  \ref{section:bmass} respectively. On the other hand, the renormalization that contains the trailing string is more complicated. We need to numerically solve the single endpoint trailing string solution for the same $(\o,LT)$ values of the corresponding rotating meson. We first determine numerically from the differential equation  \eq{rps} the constant $\Pi$ such that the spiraling string rotates around the center, for a fixed radius $LT$ with an angular frequency $\o$. The numerical relation we obtain is of the form $\Pi(LT,\o)$. Simultaneously we also identify the values $\prt{u_{ws},\r(u_{ws})}$ from the equations \eq{ucc} for the given set of values $(LT,\o)$. To find the full trailing solution of the string for each $\o$ we integrate the differential equation \eq{rps} from $u_{ws}$ to the boundary, and from $u_{ws}$ to the black hole horizon with the initial conditions obtained by solving perturbatively and analytically  the equation \eq{rps} around $u_{ws}$. We patch the two  solutions at $u_{ws}$ to get the full spiraling string. Finally to obtain the infinite energy and angular momentum of the spiralling string we numerically integrate \eq{enermom} over the obtained string worldsheet from the boundary to its causally connected $u_{ws}$ point.

\subsection{The Renormalized Energy}

In this section we present the results for the energy and the angular momentum of the observables.  The UV limit for all the renormalized observables is common and conformal while the IR differs depending on the scheme. The holographic renormalization scheme does not  depend on the state of the theory since there is no IR contribution. However for the rest of the observables the quantity that is used to renormalize the infinity does contain an extra IR term that depends on the state of the theory. This can be thought as motivated by the analogue of the thermal mass renormalization scheme for the static Wilson loop  \cite{Maldacena:1998im,Brandhuber:1998bs}.

Our theory has  two scales and the quantitative behavior of each observable with respect to the anisotropy does depend on the scheme we use to cancel the infinity. It is helpful to recall firstly the role of the different renormalization schemes in static solutions with a single scale, the temperature. Static heavy quark observables that renormalize with terms that include IR contributions, which have a typical form of $f(1/u_h)$, tend to develop total energies that are simply shifted by the IR contribution of the extra term. These terms are absent in holographic renormalization. The most characteristic example is the Wilson loop at finite temperature \cite{Brandhuber:1998bs} and anisotropy \cite{Giataganas:2012zy}. For static bound states there are two main observations we would like to focus. There exists a maximum at the renormalized finite energy, such that there are two string solutions of same length $L$, but different energy. The solution that is closer to the boundary is the stable and energetically favorable. This is a property of the solution itself and occurs irrespective of the renormalization scheme, it is evident in the holographic renormalization and mass subtraction scheme. The second observation appears in the mass renormalization scheme, when from the energy of the meson, the thermal quark masses including terms of the form $f(1/u_h)$ are subtracted. There exists a critical length (or temperature depending what dimensional quantity is kept fixed), where the energy of the two separate free quarks becomes less than the energy of the bound state. This is where the bound state prefers to dissociate. This critical length is always smaller than the length that the state becomes unstable and therefore of greater phenomenological importance.

In the case of spinning bound states the situation is more involved. We choose a string solution of boundary size $2 LT$  that is energetically favorable and belongs in the stable branch. We keep fixed the size of the meson and we change the angular velocity to observe the effect of the angular frequency and the anisotropy of the theory on the spinning heavy meson. Using the holographic renormalization we find the maximum angular frequency beyond which the bound state configuration does not exist.  While using the mass renormalization scheme of the spinning quarks we aim to compare the energies of the bound state to that of the free spinning quarks.

\begin{figure}[!t]
	\begin{minipage}[ht]{0.5\textwidth}
		\begin{flushleft}
			\centerline{\includegraphics[width=70mm]{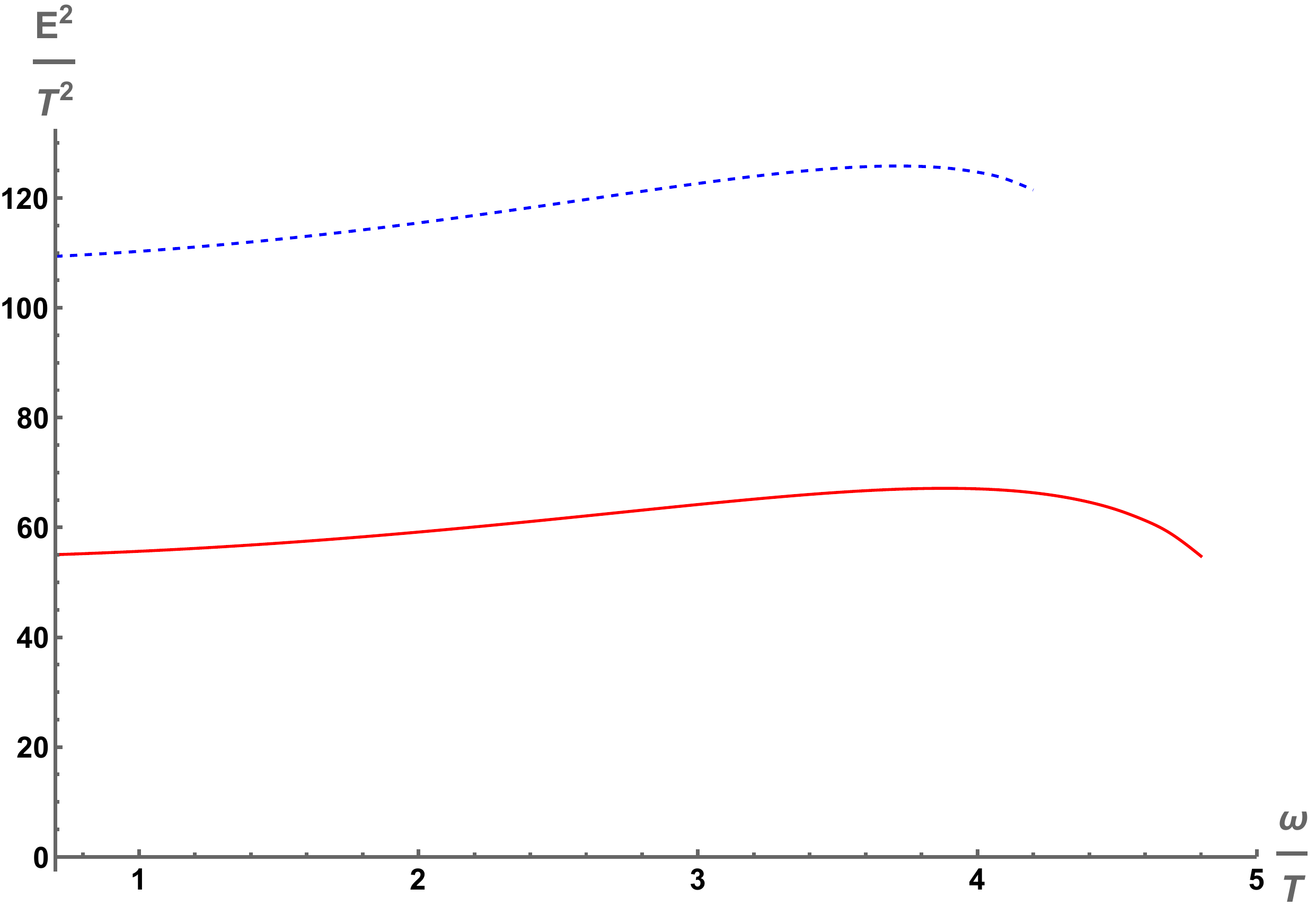}}
			\caption{\small{The energy of the spinning strings dual to rotating mesons, obtained in the holographic renormalization scheme for two different anisotropies, while keeping fixed the boundary value $LT=.1$.  We observe that increase of anisotropy leads to increase of energy. Notice the existence of an  $\o_{max}$ that the  spin can reach. This value is decreased for increasing anisotropies hinting that the range of frequencies that the bound state of meson exists decreases for higher anisotropies. The red solid curve is for $a/T\simeq0.7$ and the blue dashed for a large anisotropy $a/T\simeq8.2$.}}
			\label{figure:l1}
		\end{flushleft}
	\end{minipage}
	\hspace{0.3cm}
	\begin{minipage}[ht]{0.5\textwidth}
		\begin{flushleft}
			\centerline{\includegraphics[width=77mm ]{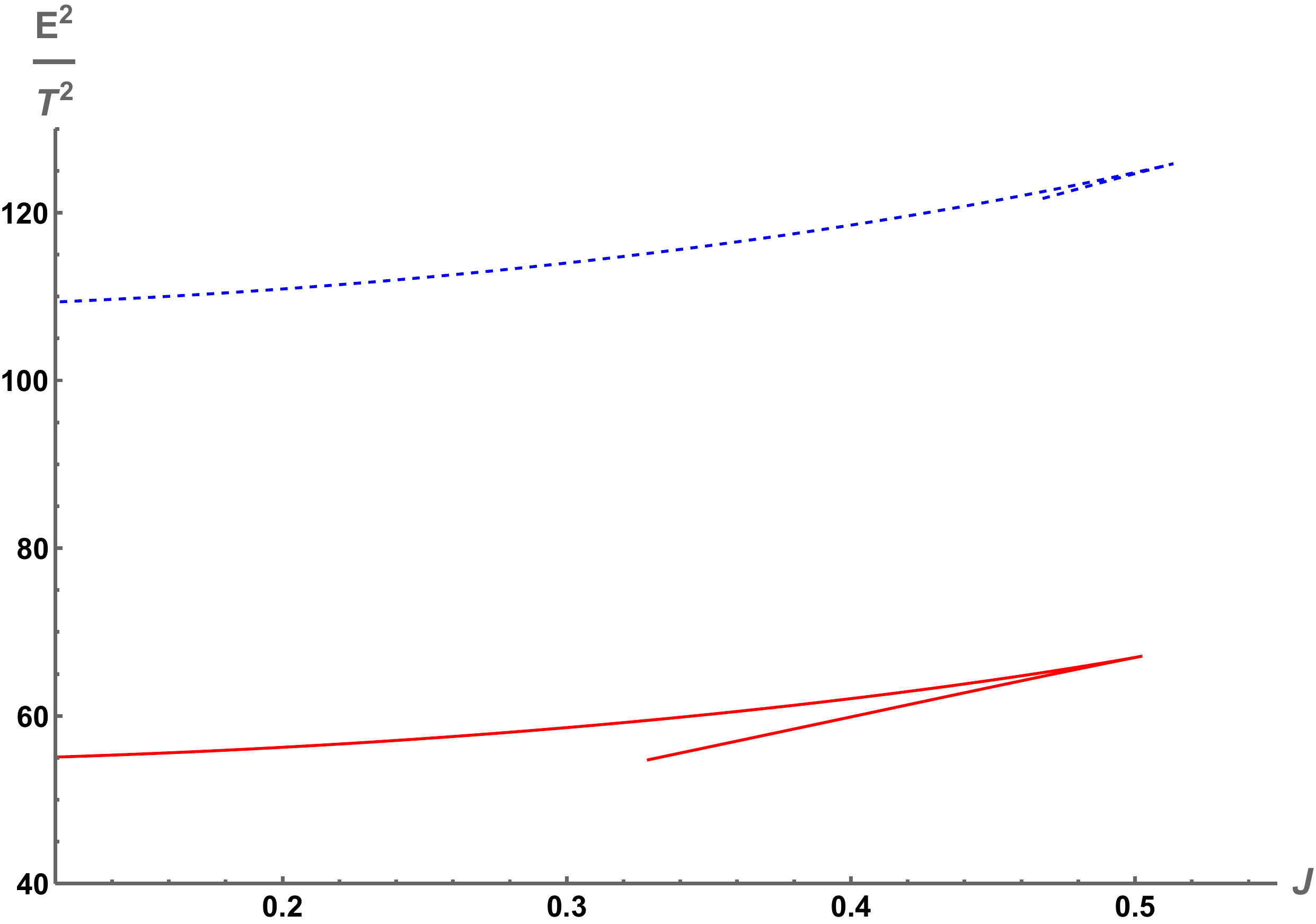}}
			\caption{\small{The energy versus the absolute angular momentum for two different rotating strings with fixed boundary size $LT$. The settings are similar to the Figure \ref{figure:l1}. There is a maximum reachable angular velocity for each particle. As we increase the angular velocity from zero, the energy increases in the upper segment of each curve, it reaches to a maximum at a certain value and then it decreases. As in the  Figure \ref{figure:l1} all the quantities are normalized with the temperature which is kept fixed and equal to unit.} }
			\label{figure:l2}\vspace{1.25cm}
		\end{flushleft}
	\end{minipage}
\end{figure}
The holographic renormalization of rotating mesons in theories of different anisotropies is depicted in Figures \ref{figure:l1} and \ref{figure:l2}. The energy is given by the equation \eq{ren1}, where the counterterms in the expressions are given by the equation \eq{ren_hol1}. The two mesons are compared for a fixed value of $L T$ for various angular frequencies in  theories with a different anisotropy. The numerical details on obtaining this plot are mentioned in subsection \eq{section:num}.  The increase of anisotropy leads to an increase of the absolute value of energy of the spinning string. We note that for the rotating bound state there exists an  $\o_{max}$  that the spin can reach. This value is decreased for increasing anisotropies implying that the range of frequencies that the bound state of meson exists decreases for higher anisotropies. This also hints that the dissociation of the heavy bound state occurs easier with increase of the anisotropy.

\begin{figure}[!t]
	\begin{minipage}[ht]{0.5\textwidth}
		\begin{flushleft}
			\centerline{\includegraphics[width=70mm]{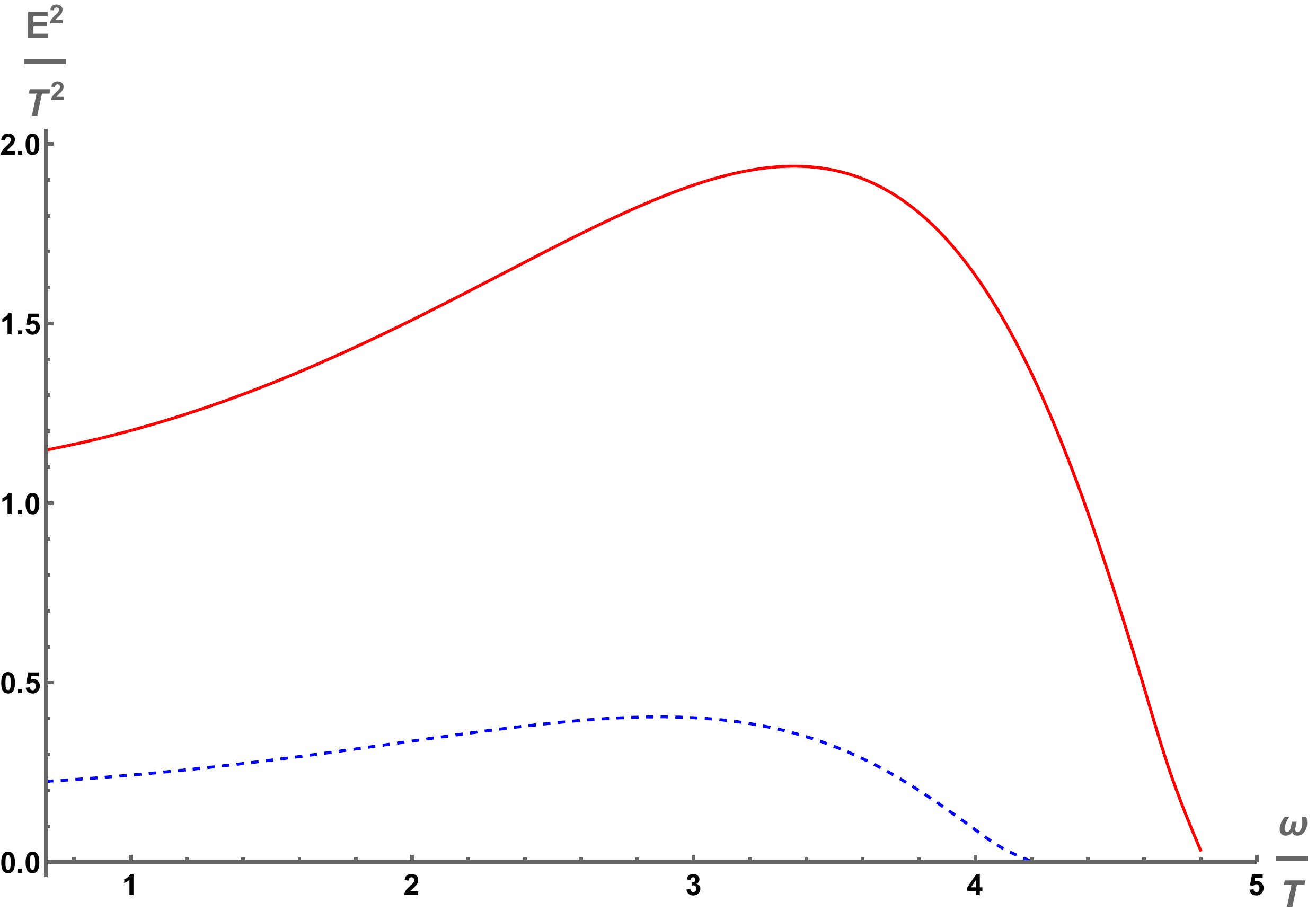}}
			\caption{\small{The energy obtained by the rotating string dual to the rotating bound state minus the approximated boosted thermal mass  \eq{ren_mass} of the quark as a function of the frequency $\o$. There exists an $\o_{max}$ that decreases as the anisotropy increases.  The plot conventions remain the same as in Figure  \ref{figure:l1}}. }
			\label{figure:d1}
		\end{flushleft}
	\end{minipage}
	\hspace{0.3cm}
	\begin{minipage}[ht]{0.5\textwidth}
		\begin{flushleft}
			\centerline{\includegraphics[width=77mm ]{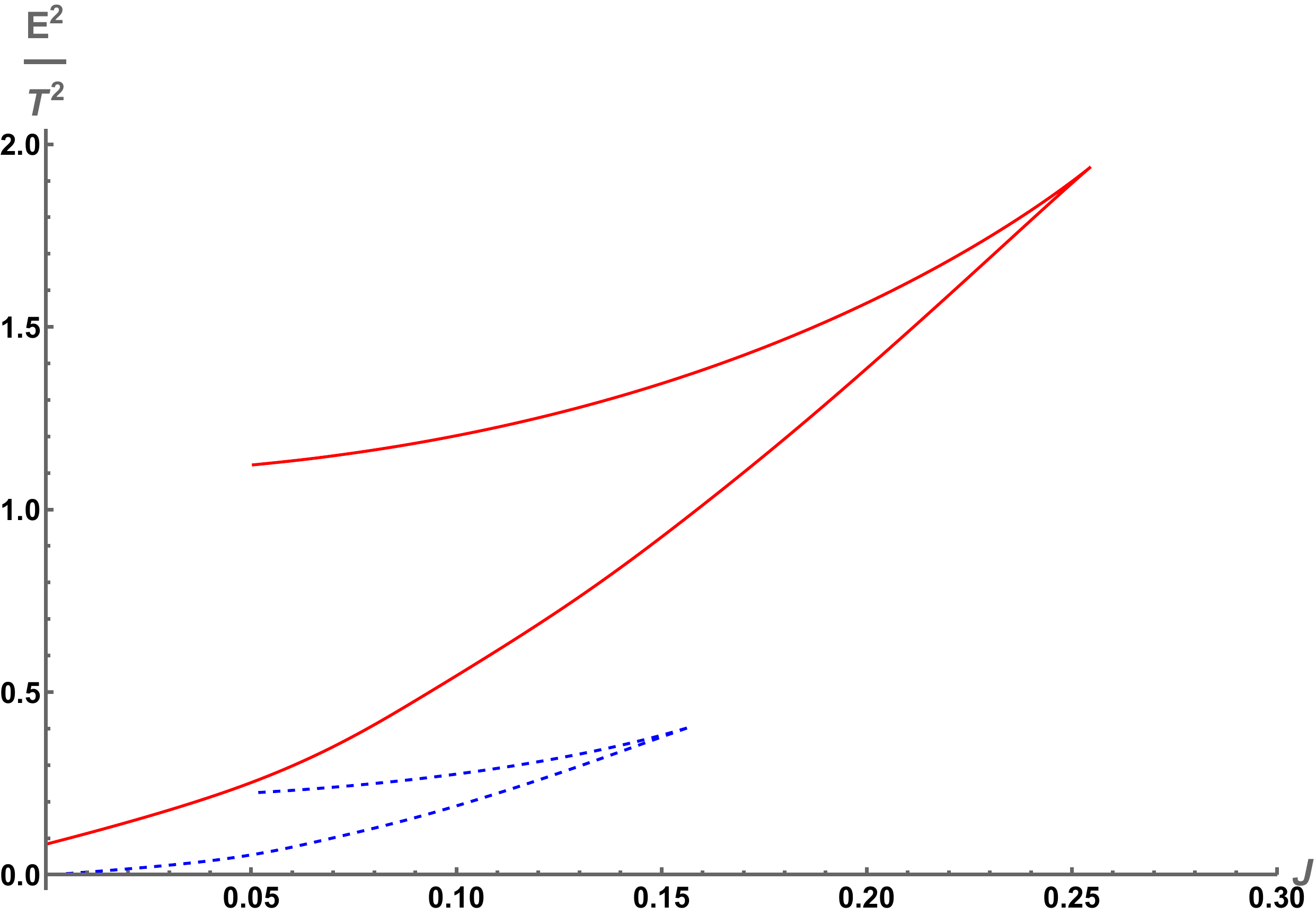}}
			\caption{\small{The absolute momentum obtained by the spiraling string dual to the rotating meson minus the approximate boosted thermal mass  \eq{ren_mass} of the quark as a function of the frequency $\o$. The plot conventions are the same as in Figure  \ref{figure:l1}.} }
			\label{figure:d2}\vspace{.8cm}
		\end{flushleft}
	\end{minipage}
\end{figure}
Next we subtract from the energy of the spinning meson the energy of the boosted thermal quark mass, introduced in subsection \ref{section:bmass}. This is an approximate scheme since the boosted quark is not an exact solution of the theory, and we neglect the drag effects. Nevertheless, at low angular frequencies the precise spiraling string solution in the thermal theories  that corresponds to the moving quark, approaches the straight string solution. Therefore, the scheme is expected to be a good and straightforward approximation at low angular velocities of the exact spinning color singlet analysis.  We will come back to this point later. We renormalize for now the energy of the meson as the equation \eq{ren_mass}, by computing the boosted thermal mass for each angular frequency and fixed $LT$.  We obtain the results plotted in Figures \ref{figure:d1} and \ref{figure:d2} for  the squared energy. Increase of the anisotropy, decreases the $\o_{max}$ as in the holographic renormalization scheme leading to the same qualitative behavior. However, we emphasize that this is an approximate scheme.

\begin{figure}[!t]
	\begin{minipage}[ht]{0.5\textwidth}
		\begin{flushleft}
			\centerline{\includegraphics[width=70mm]{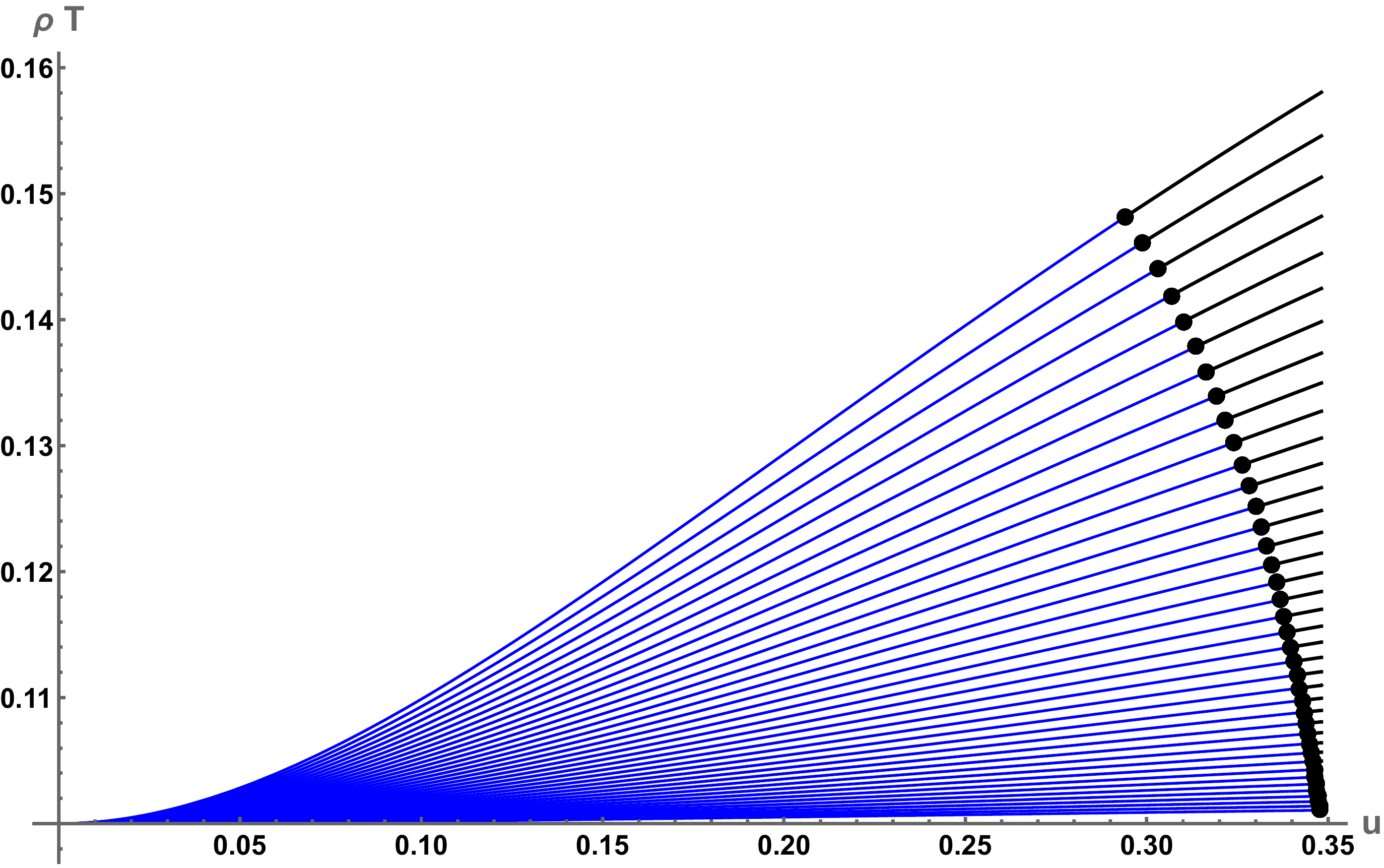}}
			\caption{\small{The profile $\r(u)$ of the trailing spiraling string for different angular frequencies $\o=(0.7,4.2)$  and a fixed radius $LT$ and anisotropy $a/T\simeq 8.2$. The lower values of $\o$ correspond to the almost straight strings, while as the frequency increases the profile of the string develops a curvature as can be seen in to upper faster spiraling strings of the plot. The worldsheet horizon is also discerned and as the frequency increases it approaches closer to the boundary of the theory. The energy of the segment of the string above the worldsheet horizon is the one we use to renormalize the energy of the meson. The boundary of the holographic background is at $u=0$. }}
			\label{figure:s1}
		\end{flushleft}
	\end{minipage}
	\hspace{0.3cm}
	\begin{minipage}[ht]{0.5\textwidth}
		\begin{flushleft}
			\centerline{\includegraphics[width=77mm ]{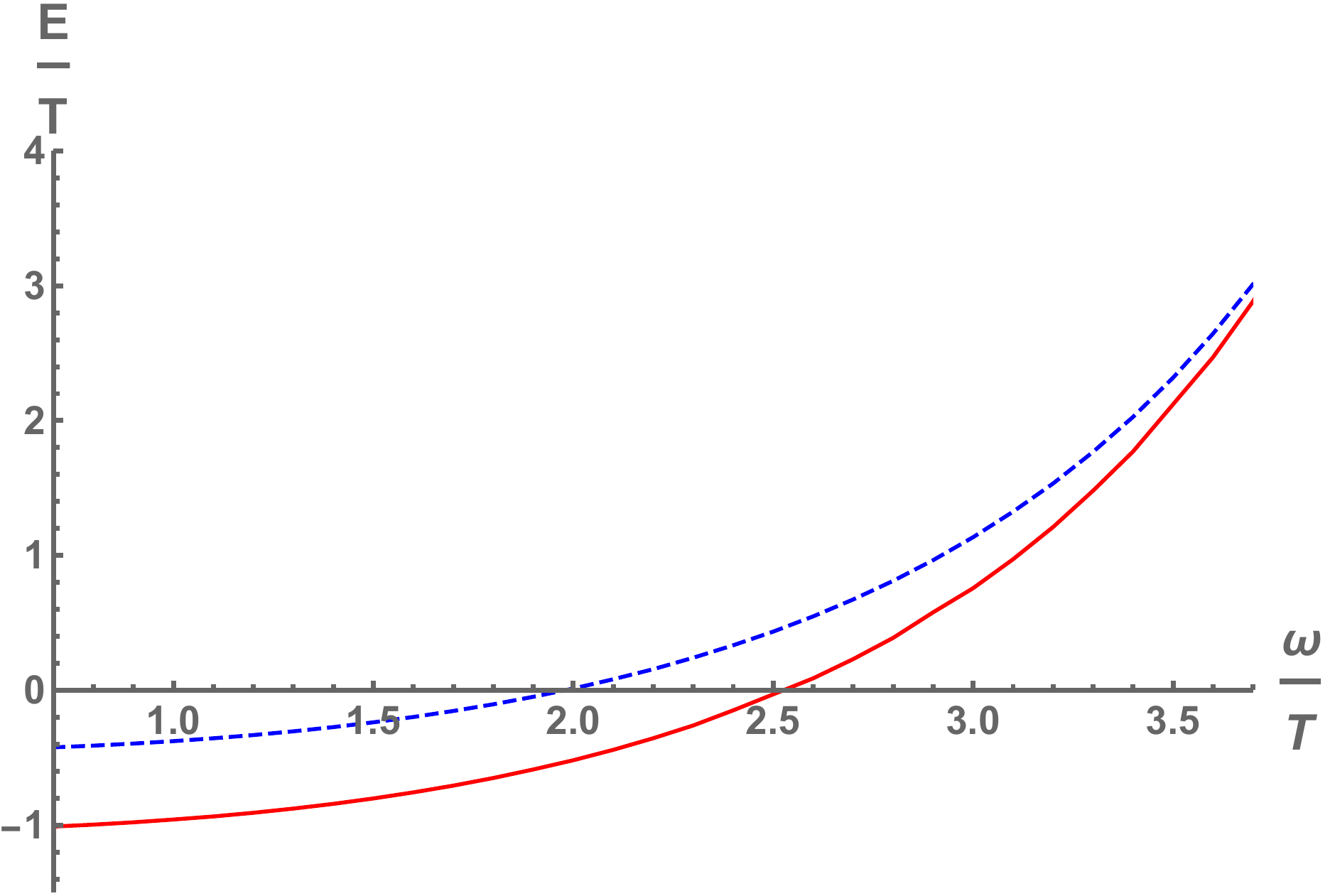}}
			\caption{\small{The total energy \eq{ren_bend12} of the rotating bound states as a function of the frequency $\o$. The energy is renormalized with the rotating color singlet energy taking into account its external source energy to maintain its motion.  The diagram is for two different anisotropies, while keeping fixed the size $LT$.  For the chosen values we observe  that there is a critical frequency such that $E_c=0$, where the bound state energy is equal to the energy of its free moving ingredients.  As the anisotropy increases the critical frequency $\o_{c}$  decreases. The conventions are the same as in Figure \ref{figure:l1}.} }
			\label{figure:s2}\vspace{1cm}
		\end{flushleft}
	\end{minipage}
\end{figure}

Lastly we proceed to the closest analogue of the mass renormalization scheme of the static mesons. For rotating mesons, it is a challenging exercise as we have elaborated already. We solve the spiraling string equations \eq{thprime1} and \eq{rps} for $\prt{\r(u),\th(u)}$,  for a fixed radius of rotation and different angular frequencies applying the methods  described in sections \ref{section:spiral} and \ref{section:num}. The projected worldsheet solution $\r(u)$ for a series of frequencies is presented in Figure \eq{figure:s1}, where the worldsheet horizon is discerned. Increase of the angular frequency, leads to increase of the drag effects and the trailing of the string, while at the same time the worldsheet horizon  moves closer to the boundary. Having obtained the string solutions we compute the energy of the string by integrating along the worldsheet from the boundary endpoint to its horizon. This choice is justified by several arguments. The boundary quark experiences an effective temperature that is given by the worldsheet horizon \cite{Giataganas:2013hwa,Gursoy:2010aa}. Moreover, fluctuations of the string outside the worldsheet horizon $u<u_{ws}$ at the boundary side, are causally disconnected from those on the other side behind the worldsheet horizon. Effectively this means that the segment of the string $u>u_{ws}$ is disconnected from its endpoint located at the boundary $u=0$, like the linear quark motion \cite{Gubser:2006bz,Casalderrey-Solana:2007ahi}. From the total energy of the spiraling string causally connected to the boundary we need to subtract the energy of the external source given by equation \eq{external}, that is provided to the color singlet to maintain its motion. Eventually we obtain the appropriate total renormalized energy \eq{ren_bend12}, which is presented in Figure \ref{figure:s2} for two mesons of the same $LT$ in different anisotropies. We observe that the energy of the bound state is lower compared to that of its free ingredients for low angular frequencies,  until a critical frequency is reached where they become equal. Increase of the theory's anisotropy leads to lower critical angular frequencies hinting an easier dissociation of the rotating meson. Here, we also note that for the study of dissociation of the mesons one may compare the actions themselves motivated by the Wilson loop expectation values. Qualitatively the same conclusions are obtained  regarding the existence of a critical frequency and its behavior.
\begin{figure}[!t]
	\begin{minipage}[ht]{0.5\textwidth}
		\begin{flushleft}
			\centerline{\includegraphics[width=70mm]{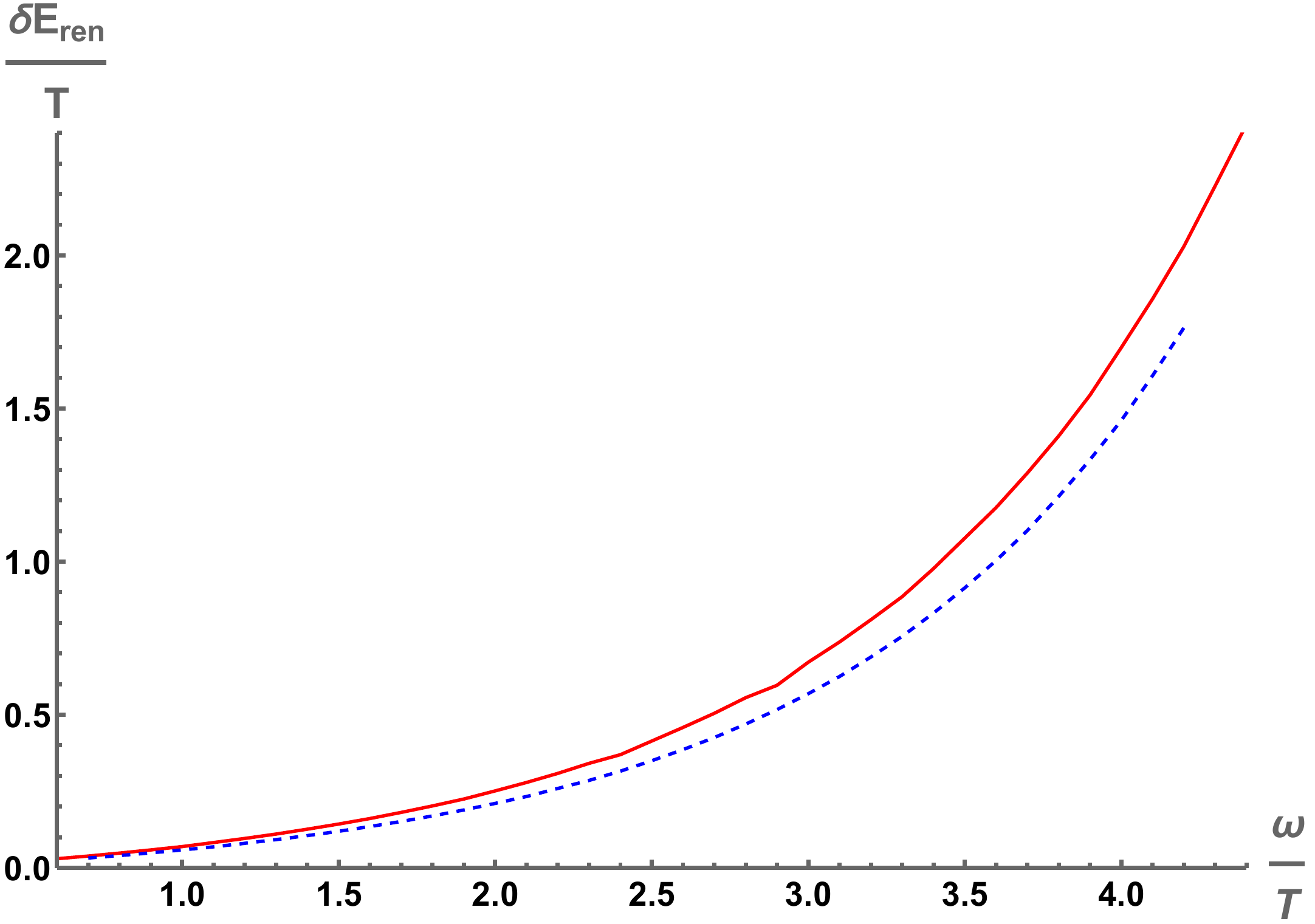}}
			\caption{\small{The difference between the energy of the spiraling string for the segment above the worldsheet horizon and the boosted static mass \eq{ren_mass}, for two different anisotropies, while keeping fixed the $LT$.  For low angular frequencies we observe the expected proximity but as the frequency increases there are notable differences.  The settings are the same as in Figure \ref{figure:l1}.}}
			\label{figure:ds1}
		\end{flushleft}
	\end{minipage}
	\hspace{0.3cm}
	\begin{minipage}[ht]{0.5\textwidth}
		\begin{flushleft}
			\centerline{\includegraphics[width=77mm ]{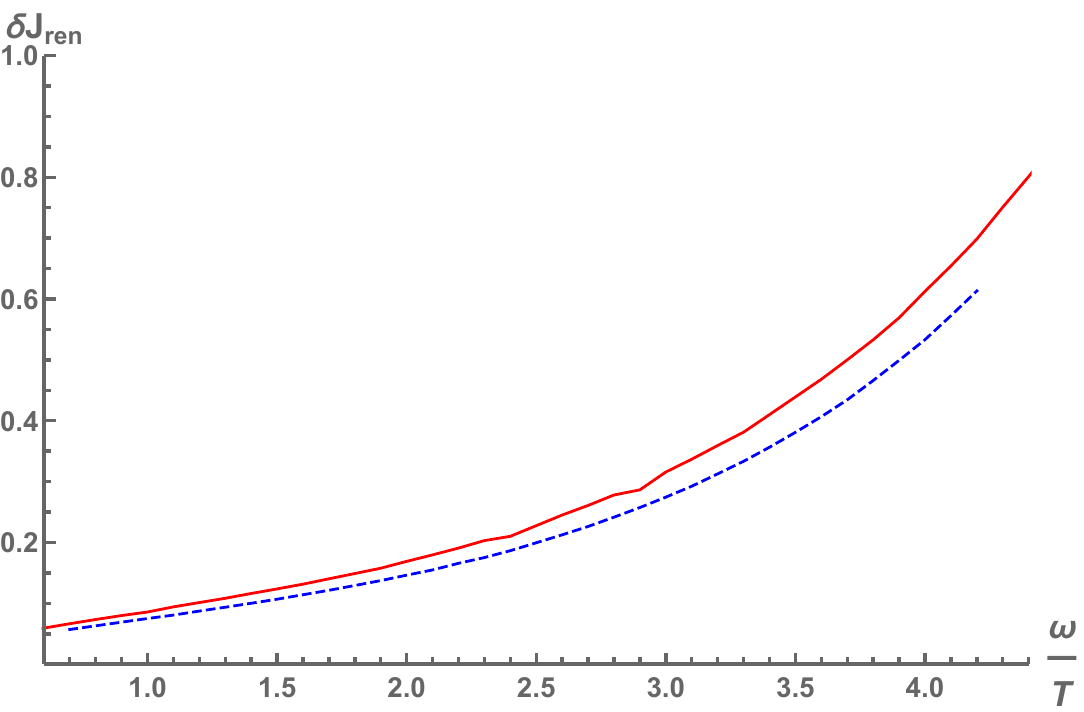}}
			\caption{\small{The difference between the angular momentum of the spiraling string for the segment above the worldsheet horizon and the angular momentum of the boosted static mass for two different anisotropies, while keeping fixed the $LT$. For higher angular frequencies the two schemes start diverging to each other.} }
			\label{figure:ds2}\vspace{.7cm}
		\end{flushleft}
	\end{minipage}
\end{figure}

Finally, motivated by the complexity of the computation of the spiraling string and the mass subtraction scheme \eq{ren_bend12}, we would like to compare it with the easily applicable approximate boosted mass scheme \eq{ren_mass}, in order to obtain the range of applicability of the latter one. We find the differences in energy and momentum for fixed $LT$ with respect to the angular frequencies, presented in Figures \ref{figure:ds1} and \ref{figure:ds2}. For low frequencies the two schemes match well, since the trailing string is almost straight and the worldsheet horizon is close to the holographic theory black hole horizon. For intermediate and higher frequencies though the schemes differ as the dragging effects become significant. As a result the boosted mass subtraction scheme may not be as reliable as the accurate mass subtraction scheme in the regime of higher angular frequencies, regarding the study of the phase transitions.

\section{Zero Temperature Spinning Strings}

Before we conclude we briefly comment on the zero temperature spinning string solutions by applying our formalism to the AdS spacetime. We find that the spinning strings in AdS are still deformed due to the angular momentum and increase of the angular frequency leads to an increase of the deformation even at zero temperature. The string solutions are qualitatively similar to the thermal case plotted earlier in Figures \ref{figure:a0} and \ref{figure:a1}.  High-energy string scattering amplitudes can be described by open classical strings ending on D-branes \cite{Gross:1987ar,Alday:2007hr}. The string spectrum at certain energies is given by semiclassical folded strings. The lowest energy rotating strings correspond to the classical limit and their analysis is included in the generic formalism we have developed here. These string solutions can be studied with the same (numerical) methods,  changing the quantities that we need fixed according to the purpose of computation; for example, we may keep fixed the position of the D-brane that the string ends instead of the string's boundary length.  Recently similar rotating string solutions have been obtained in \cite{Maldacena:2022ckr} in an effort to relate a class of deformations of the Veneziano amplitude with logarithmic Regge trajectories, the Coon amplitude, with the open string scattering amplitude for strings in AdS ending on D-branes. Both systems have similarities but also some differences at certain limits. In particular when looking at the minimum energy of a state in the Coon amplitude and comparing it to the rotating string spectrum for large momentum, there are differences in their spectrum. It may be worthy to compute the string  spectrum  when the D-brane  where the rotating string ends, is taken closer to  the boundary, so that the string deformations are taken into account in the computation of the spectrum. This is the local saddle point in the near boundary regime where the string becomes parallel to the radial direction.  Nevertheless, it is more possible that the Coon amplitudes may correspond to ordinary strings in other background than the AdS as suggested in \cite{Maldacena:2022ckr}. In this case one may find the suitable background conditions by imposing on our generic string formulas the desirable behavior and solving them for metric elements such that the lowest energy rotating string has a spectrum that agrees with the Coon amplitude behavior.

\section{Discussion}

In this work we have discussed several aspects of minimal worldsheets with rotating endpoints, corresponding to rotating heavy bound state observables.  The energy of the holographic strings, dual to heavy quarks is divergent due to the  infinite distance from the bulk to the boundary in the holographic space and therefore the infinite length of the string worldsheet. We have introduced analytically several renormalization schemes to treat the UV divergences of the bound states. They can be thought as corresponding to slightly different observables with a core the energy of the heavy meson.  This is in parallel to the energy of the static bound states. Namely we discuss the holographic renormalization \eq{ren1}, based on the expansion of the worldsheet energy in the near boundary regime, to identify the infinite term and subtract it to cancel the infinity of the energy of the bound state. We then propose the mass renormalization scheme for the rotating observables \eq{ren_bend12}, which involves the computation of the infinite energy of the two free color singlets to subtract the divergence of the energy of the heavy quark bound state energy. The renormalization schemes have the same UV counterterm contribution while differ from each other by a finite IR contribution.
Implementing the mass renormalization scheme for rotating strings is a challenging exercise. The rotating color singlets experience drag in contrast to the color neutral mesons. The drag of the quark is translated in holography to a trailing worldsheet solution that develops a  hole horizon itself. The worldsheet horizon separates the upper segment of the string ending to the boundary of the theory which moves slower than the local velocity of the light, from the lower segment of the string which ends on the horizon of the black hole spacetime and whose local velocity exceeds that of light. Moreover, the quark experiences an effective temperature determined by the worldsheet horizon.  Therefore the appropriate energy of the spiraling string we use to renormalize for our purposes is computed by integration of the string segment causally connected to the boundary from which we additionally subtract the energy of the external source that is provided to the color singlet to maintain its motion and counterbalance the drag. We have argued that the mass renormalization scheme is the appropriate scheme to study the existence of the critical frequency where the rotating bound state dissociates to its ingredients.  The rotating color singlet mass subtraction described is technically demanding. Motivated by this difficulty we also suggest an approximate scheme where we use the energy of a boosted thermal mass \eq{ren_mass} as an alternative approach to renormalize. This is applicable in a straightforward way, but it can be considered only as an approximate scheme, mostly reliable for low velocities.  The holographic framework for the renormalization schemes that we have proposed is applicable to any holographic theory.  

As a side note, we mention that one may additionally consider the application of the Legendre transform scheme \cite{Drukker:1999zq,Chu:2008xg}. The Legendre transform application lies on the fact that  the Nambu-Goto action is a functional of coordinates, while the worldsheet satisfies a Neumann boundary condition along the holographic direction at the boundary. Its application removes the UV singularity and has no IR contribution,  in this sense we expect the Legendre transform to be equivalent to the holographic renormalization scheme as in the case of static Wilson loops.

We have then applied our holographic framework  on certain thermal theories with multiple scales. We compute the non-local rotating observables in  the axion deformed thermal theories with broken rotational invariance and a non-trivial renormalization group flow with a conformal UV fixed point and a Lifshitz-like anisotropic IR fixed point.   We solve numerically the supergravity equations of motion to obtain the holographic background. Then we find numerically the minimal surface with rotating endpoints on the boundary corresponding to the heavy bound state. The rotation deforms the string and the worldsheet becomes transverse to the boundary in the bulk, before reaching the boundary which becomes transverse to boundary once more. At this point in the bulk the string distance becomes maximal in the spatial direction and we need to change the parametrization to obtain the full string solution with a patching method we apply, see for example the Figure \ref{figure:a0}. We then discuss the properties of the rotating minimal surfaces with respect to the angular frequency and anisotropy. The meson is color neutral and therefore the corresponding string is not trailing behind. The energy and momentum of the string are computed by the integration of the appropriate momenta along the string, in both parametrizations.

Then we apply the renormalization schemes of our framework. The holographic renormalization is relatively straightforward, and gives the maximum angular frequency of the meson, beyond which it cannot exist. We compare the spinning bound states of the same size, in two different anisotropies, and we find that the increase of anisotropy reduces the maximum angular frequency $\o$, hinting an easier dissociation of the bound state. Then we apply the cumbersome mass renormalization scheme to compare the energy of the bound state versus the energy of its free components undergoing the same motion. We obtain numerically the worldsheet that corresponds to the color singlet states, for the same fixed radius of spin as the bound state and the same range of angular frequencies. We shoot from the worldsheet horizon to the boundary, and separately to the black hole horizon and eventually we patch the two segments to obtain the full string solution.  We compute the energy of the string by integrating along the string from the boundary to the worldsheet horizon. The boundary quark experiences an effective temperature that is given by the worldsheet horizon, and thus this is a natural choice to make. Moreover fluctuations of the string behind the  worldsheet horizon $u<u_{ws}$ are causally disconnected from those on the other side of the horizon and this part of the string is disconnected from its endpoint located at the boundary $u=0$, justifying further this choice. From the total energy of the spiraling string causally connected to the boundary we choose to subtract the energy of the external source \eq{ren_bend12} that is provided to the color singlet to maintain its motion. Once we take all the effects into account we observe that there is a critical frequency, lower than the maximum frequency mentioned above, at which the renormalized energy becomes zero. This frequency is related qualitatively as being the one that the bound state dissociates to its ingredients. Increase of the anisotropy of the theory results to a lower critical frequency. Therefore, independent of the renormalization scheme used we find that increase of the anisotropy reduces the angular frequency that the bound state exists and the anisotropy acts as a catalyst to this type of phase transitions. This can be thought as in agreement with other qualitative behaviors of non-local observables, or even phase transitions in the strongly coupled anisotropic theories \cite{Giataganas:2012zy,Giataganas:2017koz}.

Finally we have noted that the deformation of the string worldsheet due to the angular momentum is present in zero temperature AdS spacetimes. Our framework may be useful in the recent efforts \cite{Maldacena:2022ckr} to relate the Coon amplitude and the open string scattering amplitude for strings ending on D-branes in certain backgrounds.

\textbf{Acknowledgments:}
We  would like to thank C. Hoyos  and N. Irges for useful discussions. 
The research work of D.G. is supported by the Ministry of Science and Technology of Taiwan (MOST) by the Young Scholar Columbus Fellowship grant 110-2636-M-110-008. D.G. would like to thank the Department of Theoretical Physics of CERN for hospitality during the final stages of this work.
V.G. work has been supported by the Hellenic Foundation for Research and Innovation (HFRI) and the General Secretariat for Research and Technology (GSRT), under grant agreement No 2344.

\bibliographystyle{bibhep}
%\bibliography{botanyrg}

\begin{thebibliography}{10}
	
	\bibitem{Peeters:2006iu}
	K.~Peeters, J.~Sonnenschein and M.~Zamaklar, \emph{{Holographic melting and
			related properties of mesons in a quark gluon plasma}},
	\href{https://doi.org/10.1103/PhysRevD.74.106008}{\emph{Phys. Rev.}
		{\bfseries D74} (2006) 106008},
	[\href{https://arxiv.org/abs/hep-th/0606195}{{\ttfamily hep-th/0606195}}].
	
	\bibitem{Maldacena:1998im}
	J.~M. Maldacena, \emph{{Wilson loops in large N field theories}},
	\href{https://doi.org/10.1103/PhysRevLett.80.4859}{\emph{Phys.Rev.Lett.}
		{\bfseries 80} (1998) 4859--4862},
	[\href{https://arxiv.org/abs/hep-th/9803002}{{\ttfamily hep-th/9803002}}].
	
	\bibitem{Brandhuber:1998bs}
	A.~Brandhuber, N.~Itzhaki, J.~Sonnenschein and S.~Yankielowicz, \emph{{Wilson
			loops in the large N limit at finite temperature}},
	\href{https://doi.org/10.1016/S0370-2693(98)00730-8}{\emph{Phys. Lett.}
		{\bfseries B434} (1998) 36--40},
	[\href{https://arxiv.org/abs/hep-th/9803137}{{\ttfamily hep-th/9803137}}].
	
	\bibitem{Fadafan:2008adl}
	K.~Bitaghsir~Fadafan, H.~Liu, K.~Rajagopal and U.~A. Wiedemann, \emph{{Stirring
			Strongly Coupled Plasma}},
	\href{https://doi.org/10.1140/epjc/s10052-009-0885-6}{\emph{Eur. Phys. J. C}
		{\bfseries 61} (2009) 553--567},
	[\href{https://arxiv.org/abs/0809.2869}{{\ttfamily 0809.2869}}].
	
	\bibitem{Gubser:2006bz}
	S.~S. Gubser, \emph{{Drag force in AdS/CFT}},
	\href{https://doi.org/10.1103/PhysRevD.74.126005}{\emph{Phys.Rev.} {\bfseries
			D74} (2006) 126005}, [\href{https://arxiv.org/abs/hep-th/0605182}{{\ttfamily
			hep-th/0605182}}].
	
	\bibitem{Casalderrey-Solana:2007ahi}
	J.~Casalderrey-Solana and D.~Teaney, \emph{{Transverse Momentum Broadening of a
			Fast Quark in a N=4 Yang Mills Plasma}},
	\href{https://doi.org/10.1088/1126-6708/2007/04/039}{\emph{JHEP} {\bfseries
			04} (2007) 039}, [\href{https://arxiv.org/abs/hep-th/0701123}{{\ttfamily
			hep-th/0701123}}].
	
	\bibitem{Giataganas:2013hwa}
	D.~Giataganas and H.~Soltanpanahi, \emph{{Universal Properties of the Langevin
			Diffusion Coefficients}},
	\href{https://doi.org/10.1103/PhysRevD.89.026011}{\emph{Phys.Rev.} {\bfseries
			D89} (2014) 026011}, [\href{https://arxiv.org/abs/1310.6725}{{\ttfamily
			1310.6725}}].
	
	\bibitem{Giataganas:2012zy}
	D.~Giataganas, \emph{{Probing strongly coupled anisotropic plasma}},
	\href{https://doi.org/10.1007/JHEP07(2012)031}{\emph{JHEP} {\bfseries 1207}
		(2012) 031}, [\href{https://arxiv.org/abs/1202.4436}{{\ttfamily 1202.4436}}].
	
	\bibitem{Giataganas:2018uuw}
	D.~Giataganas, \emph{{Baryons under strong magnetic fields or in theories with
			space-dependent $\theta$-term}},
	\href{https://doi.org/10.1103/PhysRevD.98.106010}{\emph{Phys. Rev. D}
		{\bfseries 98} (2018) 106010},
	[\href{https://arxiv.org/abs/1805.08245}{{\ttfamily 1805.08245}}].
	
	\bibitem{Chernicoff:2012bu}
	M.~Chernicoff, D.~Fernandez, D.~Mateos and D.~Trancanelli, \emph{{Quarkonium
			dissociation by anisotropy}},
	\href{https://doi.org/10.1007/JHEP01(2013)170}{\emph{JHEP} {\bfseries 1301}
		(2013) 170}, [\href{https://arxiv.org/abs/1208.2672}{{\ttfamily 1208.2672}}].
	
	\bibitem{Giataganas:2013zaa}
	D.~Giataganas and H.~Soltanpanahi, \emph{{Heavy Quark Diffusion in Strongly
			Coupled Anisotropic Plasmas}},
	\href{https://doi.org/10.1007/JHEP06(2014)047}{\emph{JHEP} {\bfseries 06}
		(2014) 047}, [\href{https://arxiv.org/abs/1312.7474}{{\ttfamily 1312.7474}}].
	
	\bibitem{Rajagopal:2015roa}
	K.~Rajagopal and A.~V. Sadofyev, \emph{{Chiral drag force}},
	\href{https://doi.org/10.1007/JHEP10(2015)018}{\emph{JHEP} {\bfseries 10}
		(2015) 018}, [\href{https://arxiv.org/abs/1505.07379}{{\ttfamily
			1505.07379}}].
	
	\bibitem{Giataganas:2013lga}
	D.~Giataganas, \emph{{Observables in Strongly Coupled Anisotropic Theories}},
	{\emph{PoS} {\bfseries Corfu2012} (2013) 122},
	[\href{https://arxiv.org/abs/1306.1404}{{\ttfamily 1306.1404}}].
	
	\bibitem{Giataganas:2017koz}
	D.~Giataganas, U.~G\"ursoy and J.~F. Pedraza, \emph{{Strongly-coupled
			anisotropic gauge theories and holography}},
	\href{https://doi.org/10.1103/PhysRevLett.121.121601}{\emph{Phys. Rev. Lett.}
		{\bfseries 121} (2018) 121601},
	[\href{https://arxiv.org/abs/1708.05691}{{\ttfamily 1708.05691}}].
	
	\bibitem{Giataganas:2018ekx}
	D.~Giataganas, D.-S. Lee and C.-P. Yeh, \emph{{Quantum Fluctuation and
			Dissipation in Holographic Theories: A Unifying Study Scheme}},
	\href{https://doi.org/10.1007/JHEP08(2018)110}{\emph{JHEP} {\bfseries 08}
		(2018) 110}, [\href{https://arxiv.org/abs/1802.04983}{{\ttfamily
			1802.04983}}].
	
	\bibitem{Arefeva:2020vae}
	I.~Y. Aref'eva, K.~Rannu and P.~Slepov, \emph{{Holographic model for heavy
			quarks in anisotropic hot dense QGP with external magnetic field}},
	\href{https://doi.org/10.1007/JHEP07(2021)161}{\emph{JHEP} {\bfseries 07}
		(2021) 161}, [\href{https://arxiv.org/abs/2011.07023}{{\ttfamily
			2011.07023}}].
	
	\bibitem{Gursoy:2021efc}
	U.~Gursoy, \emph{{Holographic QCD and magnetic fields}},
	\href{https://doi.org/10.1140/epja/s10050-021-00554-0}{\emph{Eur. Phys. J. A}
		{\bfseries 57} (2021) 247},
	[\href{https://arxiv.org/abs/2104.02839}{{\ttfamily 2104.02839}}].
	
	\bibitem{Arefeva:2022avn}
	I.~Y. Aref'eva, A.~Ermakov, K.~Rannu and P.~Slepov, \emph{{Holographic model
			for light quarks in anisotropic hot dense QGP with external magnetic field}},
	\href{https://arxiv.org/abs/2203.12539}{{\ttfamily 2203.12539}}.
	
	\bibitem{Golubtsova:2021agl}
	A.~A. Golubtsova, E.~Gourgoulhon and M.~K. Usova, \emph{{Heavy quarks in
			rotating plasma via holography}},
	\href{https://doi.org/10.1016/j.nuclphysb.2022.115786}{\emph{Nucl. Phys. B}
		{\bfseries 979} (2022) 115786},
	[\href{https://arxiv.org/abs/2107.11672}{{\ttfamily 2107.11672}}].
	
	\bibitem{Bohra:2020qom}
	H.~Bohra, D.~Dudal, A.~Hajilou and S.~Mahapatra, \emph{{Chiral transition in
			the probe approximation from an Einstein-Maxwell-dilaton gravity model}},
	\href{https://doi.org/10.1103/PhysRevD.103.086021}{\emph{Phys. Rev. D}
		{\bfseries 103} (2021) 086021},
	[\href{https://arxiv.org/abs/2010.04578}{{\ttfamily 2010.04578}}].
	
	\bibitem{Ipp:2020mjc}
	A.~Ipp, D.~I. M\"uller and D.~Schuh, \emph{{Anisotropic momentum broadening in
			the 2+1D Glasma: analytic weak field approximation and lattice simulations}},
	\href{https://doi.org/10.1103/PhysRevD.102.074001}{\emph{Phys. Rev. D}
		{\bfseries 102} (2020) 074001},
	[\href{https://arxiv.org/abs/2001.10001}{{\ttfamily 2001.10001}}].
	
	\bibitem{Iwasaki:2021nrz}
	S.~Iwasaki, M.~Oka and K.~Suzuki, \emph{{A review of quarkonia under strong
			magnetic fields}},
	\href{https://doi.org/10.1140/epja/s10050-021-00533-5}{\emph{Eur. Phys. J. A}
		{\bfseries 57} (2021) 222},
	[\href{https://arxiv.org/abs/2104.13990}{{\ttfamily 2104.13990}}].
	
	\bibitem{DElia:2021yvk}
	M.~D'Elia, L.~Maio, F.~Sanfilippo and A.~Stanzione, \emph{{Phase diagram of QCD
			in a magnetic background}},
	\href{https://doi.org/10.1103/PhysRevD.105.034511}{\emph{Phys. Rev. D}
		{\bfseries 105} (2022) 034511},
	[\href{https://arxiv.org/abs/2111.11237}{{\ttfamily 2111.11237}}].
	
	\bibitem{Ipp:2020nfu}
	A.~Ipp, D.~I. M\"uller and D.~Schuh, \emph{{Jet momentum broadening in the
			pre-equilibrium Glasma}},
	\href{https://doi.org/10.1016/j.physletb.2020.135810}{\emph{Phys. Lett. B}
		{\bfseries 810} (2020) 135810},
	[\href{https://arxiv.org/abs/2009.14206}{{\ttfamily 2009.14206}}].
	
	\bibitem{DElia:2021tfb}
	M.~D'Elia, L.~Maio, F.~Sanfilippo and A.~Stanzione, \emph{{Confining and chiral
			properties of QCD in extremely strong magnetic fields}},
	\href{https://doi.org/10.1103/PhysRevD.104.114512}{\emph{Phys. Rev. D}
		{\bfseries 104} (2021) 114512},
	[\href{https://arxiv.org/abs/2109.07456}{{\ttfamily 2109.07456}}].
	
	\bibitem{Fadafan:2012qu}
	K.~B. Fadafan and H.~Soltanpanahi, \emph{{Energy loss in a strongly coupled
			anisotropic plasma}},
	\href{https://doi.org/10.1007/JHEP10(2012)085}{\emph{JHEP} {\bfseries 1210}
		(2012) 085}, [\href{https://arxiv.org/abs/1206.2271}{{\ttfamily 1206.2271}}].
	
	\bibitem{Chu:2016pea}
	C.-S. Chu and D.~Giataganas, \emph{{Thermal bath in de Sitter space from
			holography}}, \href{https://doi.org/10.1103/PhysRevD.96.026023}{\emph{Phys.
			Rev. D} {\bfseries 96} (2017) 026023},
	[\href{https://arxiv.org/abs/1608.07431}{{\ttfamily 1608.07431}}].
	
	\bibitem{Gursoy:2010aa}
	U.~Gursoy, E.~Kiritsis, L.~Mazzanti and F.~Nitti, \emph{{Langevin diffusion of
			heavy quarks in non-conformal holographic backgrounds}},
	\href{https://doi.org/10.1007/JHEP12(2010)088}{\emph{JHEP} {\bfseries 1012}
		(2010) 088}, [\href{https://arxiv.org/abs/1006.3261}{{\ttfamily 1006.3261}}].
	
	\bibitem{deHaro:2000vlm}
	S.~de~Haro, S.~N. Solodukhin and K.~Skenderis, \emph{{Holographic
			reconstruction of space-time and renormalization in the AdS / CFT
			correspondence}}, \href{https://doi.org/10.1007/s002200100381}{\emph{Commun.
			Math. Phys.} {\bfseries 217} (2001) 595--622},
	[\href{https://arxiv.org/abs/hep-th/0002230}{{\ttfamily hep-th/0002230}}].
	
	\bibitem{Herzog:2006gh}
	C.~Herzog, A.~Karch, P.~Kovtun, C.~Kozcaz and L.~Yaffe, \emph{{Energy loss of a
			heavy quark moving through N=4 supersymmetric Yang-Mills plasma}},
	\href{https://doi.org/10.1088/1126-6708/2006/07/013}{\emph{JHEP} {\bfseries
			0607} (2006) 013}, [\href{https://arxiv.org/abs/hep-th/0605158}{{\ttfamily
			hep-th/0605158}}].
	
	\bibitem{Gutiez:2020sxg}
	D.~Gutiez and C.~Hoyos, \emph{{Holographic Wilsonian renormalization of a heavy
			quark moving through a strongly coupled plasma}},
	\href{https://doi.org/10.1007/JHEP10(2020)119}{\emph{JHEP} {\bfseries 10}
		(2020) 119}, [\href{https://arxiv.org/abs/2007.11284}{{\ttfamily
			2007.11284}}].
	
	\bibitem{Azeyanagi:2009pr}
	T.~Azeyanagi, W.~Li and T.~Takayanagi, \emph{{On String Theory Duals of
			Lifshitz-like Fixed Points}},
	\href{https://doi.org/10.1088/1126-6708/2009/06/084}{\emph{JHEP} {\bfseries
			0906} (2009) 084}, [\href{https://arxiv.org/abs/0905.0688}{{\ttfamily
			0905.0688}}].
	
	\bibitem{Mateos:2011ix}
	D.~Mateos and D.~Trancanelli, \emph{{The anisotropic N=4 super Yang-Mills
			plasma and its instabilities}},
	\href{https://doi.org/10.1103/PhysRevLett.107.101601}{\emph{Phys.Rev.Lett.}
		{\bfseries 107} (2011) 101601},
	[\href{https://arxiv.org/abs/1105.3472}{{\ttfamily 1105.3472}}].
	
	\bibitem{Giataganas:2018ekxv}
	D.~Giataganas, D.-S. Lee and C.-P. Yeh, \emph{Vacuum solution of section 7.1.2,
		{Quantum Fluctuation and Dissipation in Holographic Theories: A Unifying
			Study Scheme}}, \href{https://doi.org/10.1007/JHEP08(2018)110}{\emph{JHEP}
		{\bfseries 08} (2018) 110},
	[\href{https://arxiv.org/abs/1802.04983}{{\ttfamily 1802.04983}}].
	
	\bibitem{Inkof:2019gmh}
	G.~A. Inkof, J.~M.~C. K\"uppers, J.~M. Link, B.~Gout\'eraux and J.~Schmalian,
	\emph{{Quantum critical scaling and holographic bound for transport
			coefficients near Lifshitz points}},
	\href{https://doi.org/10.1007/JHEP11(2020)088}{\emph{JHEP} {\bfseries 11}
		(2020) 088}, [\href{https://arxiv.org/abs/1907.05744}{{\ttfamily
			1907.05744}}].
	
	\bibitem{Gross:1987ar}
	D.~J. Gross and P.~F. Mende, \emph{{String Theory Beyond the Planck Scale}},
	\href{https://doi.org/10.1016/0550-3213(88)90390-2}{\emph{Nucl. Phys. B}
		{\bfseries 303} (1988) 407--454}.
	
	\bibitem{Alday:2007hr}
	L.~F. Alday and J.~M. Maldacena, \emph{{Gluon scattering amplitudes at strong
			coupling}}, \href{https://doi.org/10.1088/1126-6708/2007/06/064}{\emph{JHEP}
		{\bfseries 06} (2007) 064},
	[\href{https://arxiv.org/abs/0705.0303}{{\ttfamily 0705.0303}}].
	
	\bibitem{Maldacena:2022ckr}
	J.~Maldacena and G.~N. Remmen, \emph{{Accumulation-Point Amplitudes in String
			Theory}},  \href{https://arxiv.org/abs/2207.06426}{{\ttfamily 2207.06426}}.
	
	\bibitem{Drukker:1999zq}
	N.~Drukker, D.~J. Gross and H.~Ooguri, \emph{{Wilson loops and minimal
			surfaces}}, \href{https://doi.org/10.1103/PhysRevD.60.125006}{\emph{Phys.
			Rev.} {\bfseries D60} (1999) 125006},
	[\href{https://arxiv.org/abs/hep-th/9904191}{{\ttfamily hep-th/9904191}}].
	
	\bibitem{Chu:2008xg}
	C.-S. Chu and D.~Giataganas, \emph{{UV-divergences of Wilson Loops for
			Gauge/Gravity Duality}},
	\href{https://doi.org/10.1088/1126-6708/2008/12/103}{\emph{JHEP} {\bfseries
			0812} (2008) 103}, [\href{https://arxiv.org/abs/0810.5729}{{\ttfamily
			0810.5729}}].
	
\end{thebibliography}
\providecommand{\href}[2]{#2}\begingroup\raggedright
\end{document}